\documentclass[aps,twocolumn,prd,showpacs,showkeys,preprintnumbers,superscriptaddress,nobibnotes,floatfix,longbibliography,nofootinbib]{revtex4-2}
\pdfoutput=1
\usepackage{amsmath}
\usepackage{amsfonts}
\usepackage{amssymb}
\usepackage{mathrsfs}
\usepackage{graphicx}
\usepackage{color}
\usepackage{hyperref}
\usepackage{multirow}
\usepackage{slashed}
\usepackage{epsf}
\usepackage{cancel}

\usepackage{graphicx}
\usepackage{caption}
\usepackage{subcaption}

\usepackage{color}

\def\to{\rightarrow}

\begin{document}

\title{ LHC Run-3, Dark Matter and Supersymmetric Spectra in the Spersymmetric Pati-Salam Model}

\author{Ali Muhammad}
\affiliation{CAS Key Laboratory of Theoretical Physics, Institute of Theoretical Physics, Chinese Academy of Sciences, Beijing 100190, China}
\affiliation{School of Physical Sciences, University of Chinese Academy of Sciences, No. 19A Yuquan Road, Beijing 100049, China}

\author{Imtiaz Khan}
\email{ikhanphys1993@gmail.com}
\affiliation{Department of Physics, Zhejiang Normal University, Jinhua, Zhejiang 321004, China}
\affiliation{Research Center of Astrophysics and Cosmology, Khazar University, Baku, AZ1096, 41 Mehseti Street, Azerbaijan}
\affiliation{Zhejiang Institute of Photoelectronics, Jinhua, Zhejiang 321004, China}

\author{Tianjun Li}

\affiliation{School of Physics, Henan Normal University, Xinxiang 453007, P. R. China}

\author{Shabbar Raza}
\affiliation{Department of Physics, Federal Urdu University of Arts, Science and Technology, Karachi 75300, Pakistan}

\author{Mussawir Khan} 
\affiliation{State Key Laboratory of Particle Astrophysics, Institute of High Energy Physics, Chinese Academy of Sciences, Beijing 100049, China}
\affiliation{University of Chinese Academy of Sciences, Beijing 100049, China}

\author{Pirzada 
 }%
\affiliation{CAS Key Laboratory of Theoretical Physics, Institute of Theoretical Physics, Chinese Academy of Sciences, Beijing 100190, China}
\affiliation{School of Physical Sciences, University of Chinese Academy of Sciences, No. 19A Yuquan Road, Beijing 100049, China}

\begin{abstract}
We reexamine phenomenological consequences of the Minimal Supersymmetric Standard Model (MSSM) 
arising from the supersymmetric $SU(4)_C \times SU(2)_L \times SU(2)_R$ Pati-Salam model, considering both positive and negative values of the Higgsino mass parameter $\mu$, as the muon anomalous magnetic moment might be in accordance with the Standard Model (SM) prediction. We consider the current constraints from the LHC supersymmetry searches, precision Higgs measurements, and direct dark matter detection experiments. We identify the viable parameter spaces in which the observed dark matter relic abundance, as reported by Planck~2018, is achieved through several mechanisms, including gluino-neutralino, sbottom-neutralino, stop-neutralino, stau-neutralino, and chargino-neutralino coannihilations, as well as resonant annihilation via the pseudoscalar Higgs boson ($A$-funnel). We demonstrate that all these scenarios are still consistent with the existing collider bounds and current limits on spin-independent and spin-dependent neutralino-nucleon scattering cross sections. In particular, we find that sbottom-neutralino coannihilation solutions typically require sbottom masses near $2.8~\text{TeV}$, while gluino-neutralino and stop-neutralino coannihilation scenarios allow gluino masses in the range $1$--$3~\text{TeV}$ and stop masses between $1$ and $3.5~\text{TeV}$. In coannihilation-dominated regions, the stau and chargino masses can reach values as high as $3.5~\text{TeV}$, whereas viable $A$-resonance solutions are found for pseudoscalar Higgs masses spanning approximately $1.7$--$3.8~\text{TeV}$. We anticipate that a significant portion of the surviving parameter space will be accessible to the supersymmetry searches at the LHC Run 3 and future upgrades. Most of these scenarios remain within the reach of current and upcoming direct-detection experiments, such as XENONnT, LZ current, and LZ 1000-day sensitivity.
\end{abstract}

\maketitle


\section{Introduction}
\label{intro}
Among theories for the Standard Model (SM) extension, supersymmetry (SUSY) offers a highly motivated approach to physics at energy scales beyond electroweak symmetry breaking. These theories naturally achieve gauge coupling unification at high energies~\cite{gaugeunification}, protect the Higgs mass from destabilizing quantum corrections~\cite{Witten:1981nf}, and, under $R$ parity conservation yield a stable weakly interacting massive particle capable of explaining cosmological dark matter observations~\cite{darkmatterreviews}. The Minimal Supersymmetric Standard Model (MSSM) specifically establishes an upper limit for the lightest CP-even Higgs boson mass $m_h \lesssim 135~\text{GeV}$~\cite{Slavich:2020zjv}. This prediction aligns with the LHC discovery of a Higgs boson at approximately $125~\text{GeV}$~\cite{ATLAS:2012yve,CMS:2012qbp}. This experimental result has significantly sharpened the viable parameter space of supersymmetric models and motivates detailed phenomenological studies of well-motivated SUSY constructions.

Unified supersymmetric frameworks incorporating grand gauge symmetries such as $SU(5)$ and $SO(10)$ SUSY GUT, or the left-right symmetric $SU(4)_C \times SU(2)_L \times SU(2)_R$ Pati-Salam model provide the notable advantage of enabling the third-generation Yukawa coupling unification. The unification of the top, bottom, and tau Yukawa couplings ($t$--$b$--$\tau$) in the $SO(10)$ and Pati-Salam model, orthe $b$--$\tau$ unification in the $SU(5)$ model can be realized at the grand unification scale~\cite{big-422,bigger-422,Baer:2008jn,Gogoladze:2009ug,Baer:2009ff,Gogoladze:2009bn,Gogoladze:2010fu}. Recent developments in both supersymmetric and non-supersymmetric settings further highlight the continued relevance of Yukawa unification as a probe of high-scale physics~\cite{Gomez:2020gav,Djouadi:2022gws}.
A distinctive feature of the supersymmetric Pati-Salam model is the non-universality of gaugino masses induced by the underlying gauge structure. At the grand unification boundary, the soft supersymmetry-breaking parameters for the $U(1)_Y$, $SU(2)_L$, and $SU(3)_c$ gauge sectors obey a specific mass relation dictated by the symmetry breaking pattern
\begin{equation}
M_1 = \frac{3}{5} M_2 + \frac{2}{5} M_3,
\end{equation}
which leads to phenomenologically rich low-energy spectra. This non-universality in the gaugino masses, together with the sign of the Higgsino mass parameter $\mu$, plays a crucial role in shaping the viable parameter space of the model. In particular, it has been demonstrated that $\mu < 0$ yields favorable finite threshold corrections to the bottom quark Yukawa coupling, enabling successful $t$--$b$--$\tau$ Yukawa unification while maintaining a relatively light superpartner spectrum~\cite{Gogoladze:2010fu}.
Remarkably, the SUSY Pati-Salam framework uniquely accommodates gluino--neutralino coannihilation scenarios that simultaneously satisfy the observed dark matter relic density and achieve $t$--$b$--$\tau$ Yukawa unification at the level of $10\%$ or better~\cite{Gogoladze:2009ug,Gogoladze:2009bn,Profumo:2004wk,Ajaib1}. This distinctive interplay between collider constraints, flavor physics, and dark matter phenomenology renders the SUSY Pati-Salam model a particularly attractive target for exploration in light of current and upcoming LHC data.
Previous studies have demonstrated that successful $t$--$b$--$\tau$ Yukawa unification within the SUSY Pati-Salam framework can be achieved when the soft supersymmetry-breaking gaugino masses share a common sign, while simultaneously yielding a neutralino lightest supersymmetric particle (LSP) consistent with the observed dark matter relic abundance via gluino--neutralino coannihilation~\cite{Gogoladze:2009ug,Gogoladze:2009bn,Profumo:2004wk}. However, allowing for non-trivial sign assignments among the gaugino mass parameters significantly enlarges the viable parameter space. In particular, for $\mu < 0$ with $M_{2} < 0$ and $M_{3} > 0$, phenomenologically acceptable solutions satisfying the current experimental constraints and realizing $t$--$b$--$\tau$ Yukawa unification at the level of $10\%$ or better are obtained for scalar soft masses as low as $m_{0} \gtrsim 300~\text{GeV}$. This represents a substantial reduction compared to scenarios with same-sign gaugino masses, which typically require $m_{0} \gtrsim 8~\text{TeV}$, where $m_{0}$ denotes the universal scalar soft mass parameter at the grand unification scale $M_{\rm GUT}$~\cite{Gogoladze:2010fu}.
In this mixed-sign gaugino configuration, a variety of dark matter annihilation mechanisms become operative. In addition to gluino--neutralino coannihilation, viable relic density can be achieved through chargino--neutralino and stau--neutralino coannihilation channels, as well as through resonant annihilation mediated by the pseudoscalar Higgs boson (the $A$-funnel)~\cite{Gogoladze:2010fu,Gogoladze:2012ii}. These complementary mechanisms highlight the rich dark matter phenomenology inherent in the SUSY Pati-Salam model.
It is important to emphasize that while exact $t$--$b$--$\tau$ Yukawa unification arises naturally in the minimal Pati-Salam construction, it needs not be preserved once higher-dimensional operators are taken into account. Such operators can modify the boundary conditions for the Yukawa couplings at $M_{\rm GUT}$, allowing for scenarios in which $b$--$\tau$ unification is maintained while full $t$--$b$--$\tau$ unification is relaxed. In particular, one may consider classes of higher-dimensional contributions that yield $y_{e}/y_{d} = 1$ but $y_{u}/y_{d} \neq 1$, thereby preserving $b$--$\tau$ unification within 
the Pati-Salam framework while departing from exact third-generation Yukawa unification~\cite{Antusch:2013rxa,Antusch:2009gu,Trine:2009ns}.

The interplay of the non-universal gaugino masses with the sign of the Higgsino mass parameter $\mu$ generates distinctive and complex phenomenology in supersymmetric $SU(4)_C \times SU(2)_L \times SU(2)_R$ Pati-Salam models. Crucially, the sign of $\mu$ critically shapes low-energy observables by governing radiative corrections to precision electroweak parameters, influencing flavor-changing processes, and determining dark matter relic density predictions. As shown in Ref.~\cite{Gogoladze:2010fu,Ahmed:2022ibc}, choosing $\mu < 0$ is essential for obtaining the appropriate finite threshold corrections to the bottom quark and bottom $\tau$ Yukawa coupling, respectively, thereby enabling successful third-generation Yukawa unification ($t$--$b$--$\tau$) while maintaining a comparatively light supersymmetric spectrum. In this regime, the predicted bottom-quark mass is in better agreement with experimental data without requiring excessively heavy scalar masses, making $\mu < 0$ particularly attractive from a unification standpoint.
The Higgsino mass parameter's sign $\mu$ significantly influences predictions for the muon's anomalous magnetic moment $(g-2)_\mu$. Within supersymmetric frameworks, the dominant corrections to this observable stem from one-loop processes involving virtual exchanges between smuons and neutralinos together with sneutrino-chargino interactions. These supersymmetric contributions can be expressed through parametric approximations as demonstrated in~\cite{KhalilS2017}
\begin{equation}
\Delta a_\mu^{\rm SUSY} \sim \frac{M_i \, \mu \, \tan\beta}{m_{\rm SUSY}^4},
\label{eq:g-2}
\end{equation}
where $M_1$ and $M_2$ denote the bino and wino masses, respectively, $\tan\beta = \langle H_u \rangle / \langle H_d \rangle$ represents the ratio of neutral Higgs vacuum expectation values, and $m_{\rm SUSY}$ characterizes the dominant superpartner mass scale contributing to the loop diagrams. Eq.~\eqref{eq:g-2} makes it explicit that the magnitude and sign of the supersymmetric contributions are directly correlated with the sign of $\mu$. Historically, theorists have favored $\mu>0$ because it generates positive contributions to the muon anomalous magnetic moment, which were essential for reconciling the formerly observed discrepancy between the SM predictions and experimental measurements of $(g-2)_\mu$~\cite{Ahmed:2021htr}.
This landscape has shifted dramatically due to the Fermilab Muon  $(g-2)$ Collaboration's successive high-precision results (E989), which through combined analyses of Run-1 to Run-3 data (2020–2023) have ushered in a new precision era for the muon anomalous magnetic moment measurement~\cite{Muong-2:2023cdq,Muong-2:2021ojo,Muong-2:2025xyk}. The updated world-average experimental value is now given by $a_\mu^{\rm exp} = 116\,592\,0715(145) \times 10^{-12}$~\cite{Muong-2:2025xyk},
representing an unprecedented level of experimental precision. In parallel, substantial progresses in lattice QCD calculations of the leading-order hadronic vacuum polarization contribution have led to a refined SM prediction $a_\mu^{\rm SM} = 116\,592\,033(62) \times 10^{-11}$~\cite{Aliberti:2025beg}.
As a consequence, the updated difference between experiment and theory, $\Delta a_\mu = a_\mu^{\rm exp} - a_\mu^{\rm SM} = 38.5(63.7) \times 10^{-11}$,
corresponds to a deviation of only $0.6\sigma$, effectively relaxing the previously reported $4.2\sigma$ tension. In light of this development, scenarios with $\mu < 0$ regain strong phenomenological relevance, as they are no longer disfavored by the muon anomalous magnetic moment constraint.
Beyond $(g-2)_\mu$, the sign of $\mu$ critically affects dark matter phenomenology, particularly in realizing Higgs- and $Z$-mediated annihilation channels for a neutralino lightest supersymmetric particle. These channels are subject to stringent constraints from the LHC electroweakino searches and dark matter direct-detection experiments such as LZ~\cite{Khan:2025azf,Barman:2022jdg}. Taken together, these considerations strongly motivate renewed phenomenological investigations of supersymmetric grand unified theories in the $\mu < 0$ regime.
With the onset of the LHC Run-3 and a new phase of data taking, a comprehensive reassessment of the supersymmetric parameter spaces for both signs of $\mu$ is timely and well motivated. In this work, we perform a detailed phenomenological analysis of the supersymmetric
Pati-Salam model, considering both $\mu > 0$ and $\mu < 0$ scenarios. Our study incorporates the current LHC constraints, Higgs mass measurements, and bounds from direct dark matter detection experiments. We identify the viable parameter spaces exhibiting the distinctive sparticle spectra and multiple mechanisms for reproducing the observed dark matter relic density, including $A/H$-resonance annihilation and coannihilation channels, for example, chargino-neutralino, stau-neutralino, stop-neutralino, sbottom-neutralino, and gluino-neutralino coannihilations. We demonstrate compatibility of these solutions with the recent LHC SUSY searches (including Run-3 data) and future experimental projections. Additionally, our solutions comply with the Planck 2018 relic density constraints, and numerous scenarios remain testable by the current and next-generation direct detection experiments like XENONnT and LZ~\cite{LZ:2022lsv,LZ:2018qzl,XENON:2023cxc,LZ:2024zvo}, establishing the SUSY Pati-Salam framework a compelling target for both collider and astroparticle searches.

\section{Model Parameters, Scanning Strategy, and Experimental Constraints}
\label{sec:scan}

We begin by specifying the set of fundamental parameters defining the supersymmetric $SU(4)_C \times SU(2)_L \times SU(2)_R$ Pati-Salam model. At the grand unification scale $M_{\rm GUT}$, the model is characterized by the following soft supersymmetry-breaking (SSB) parameters:
\begin{align}
m_{0}, \; m_{H_u}, \; m_{H_d}, \; A_0, \; M_2, \; M_3, \; \tan\beta, \; {\rm sign}(\mu).
\label{params}
\end{align}
Here, $m_0$ denotes the universal soft supersymmetry-breaking sfermion mass, while $m_{H_u}$ and $m_{H_d}$ represent the soft mass parameters of the up-type and down-type. Higgs doublets, respectively, and $A_0$ is the universal trilinear coupling. The parameters $M_2$ and $M_3$ correspond to the $SU(2)_L$ and $SU(3)_c$ gaugino masses. All parameters, except for $\tan\beta$ and the sign of $\mu$, are defined at $M_{\rm GUT}$; $\tan\beta \equiv v_u / v_d$ and the sign of the Higgsino mass parameter, $\mu$ are defined at the electroweak scale.

To explore the viable parameter space, we perform random scans using the \texttt{ISAJET}~7.85 package~\cite{ISAJET}. Gauge coupling unification is imposed by requiring $g_1 = g_2 = g_U$ at $M_{\rm GUT}$, while allowing the strong coupling $g_3$ to deviate by up to $3\%$ to account for unknown threshold corrections at the grand unification scale~\cite{Hisano:1992jj}. Further details of the numerical procedure and the implementation of renormalization group evolution in \texttt{ISAJET} can be found in Refs.~\cite{ISAJET,bartol2}.

The scanned ranges for the input parameters are chosen as follows:
\begin{gather}
0~{\rm TeV} \leq m_{0}, \; m_{H_u}, \; m_{H_d} \leq 20~{\rm TeV}, \nonumber\\
-10~{\rm TeV} \leq M_2 \leq 10~{\rm TeV}, \nonumber\\
0~{\rm TeV} \leq M_3 \leq 5~{\rm TeV}, \nonumber\\
30 \leq \tan\beta \leq 55, \nonumber\\
-3 \leq A_0/m_0 \leq 3, \nonumber\\
\mu > 0, {\rm and} \,\, \mu < 0.
\label{parameterRange}
\end{gather}

The parameter space is explored using the Metropolis-Hastings Markov Chain Monte Carlo algorithm~\cite{Belanger:2009ti}. We retain only points consistent with radiative electroweak symmetry breaking (REWSB) and a neutralino lightest supersymmetric particle (LSP), thereby excluding scenarios with stable charged relics~\cite{Beringer:1900zz}. We further impose current experimental bounds on sparticle masses~\cite{Agashe:2014kda}, as well as constraints from flavor observables and rare decays, including $B_s \rightarrow \mu^+ \mu^-$~\cite{Aaij:2012nna}, $b \rightarrow s \gamma$~\cite{Amhis:2012bh}, and $B_u \rightarrow \tau \nu_\tau$~\cite{Asner:2010qj}. Constraints from LHC searches for gluinos and first- and second-generation squarks are also applied~\cite{Vami:2019slp}. Finally, the neutralino relic density is required to satisfy the Planck 2018 $5\sigma$ bounds~\cite{Ade:2015xua}.
Explicitly, the applied constraints are:
\begin{gather}
122~{\rm GeV} \leq m_h \leq 128~{\rm GeV}, \\
m_{\tilde{g}} \geq 2.3~{\rm TeV}, \qquad m_{\tilde{q}} \geq 2.0~{\rm TeV}, \\
0.8 \times 10^{-9} \leq {\rm BR}(B_s \rightarrow \mu^+ \mu^-) \leq 6.2 \times 10^{-9} \; (2\sigma), \\
2.99 \times 10^{-4} \leq {\rm BR}(b \rightarrow s \gamma) \leq 3.87 \times 10^{-4} \; (2\sigma), \\
0.15 \leq \frac{{\rm BR}(B_u \rightarrow \tau \nu_\tau)_{\rm MSSM}}{{\rm BR}(B_u \rightarrow \tau \nu_\tau)_{\rm SM}} \leq 2.41 \; (3\sigma), \\
0.114 \leq \Omega_{\rm CDM} h^2 \leq 0.126 \; (5\sigma).
\end{gather}
\section{Numerical Results and Discussions}

\begin{figure*}[th!]
	\centering \includegraphics[width=8.90cm]{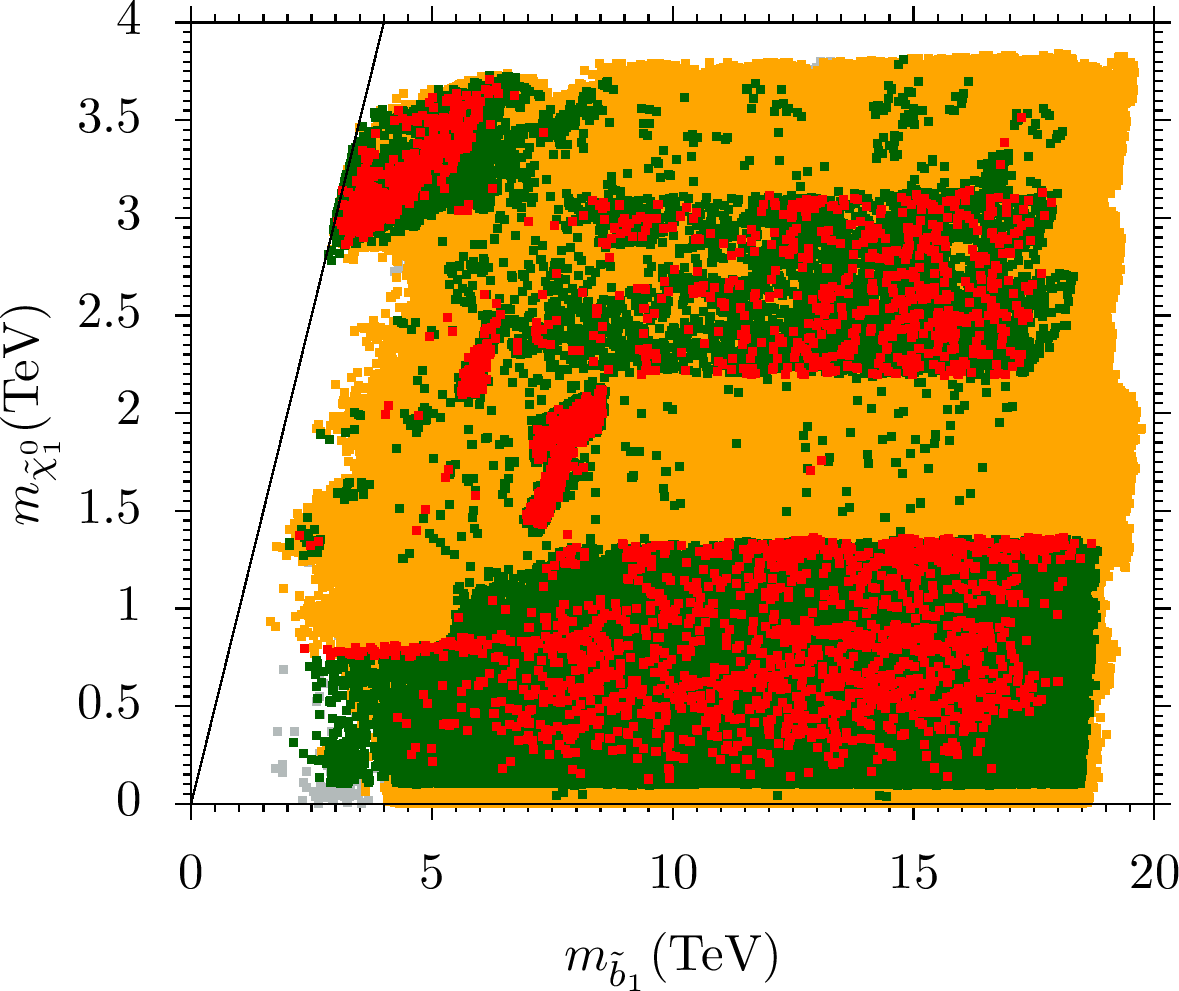}
    \centering \includegraphics[width=8.90cm]{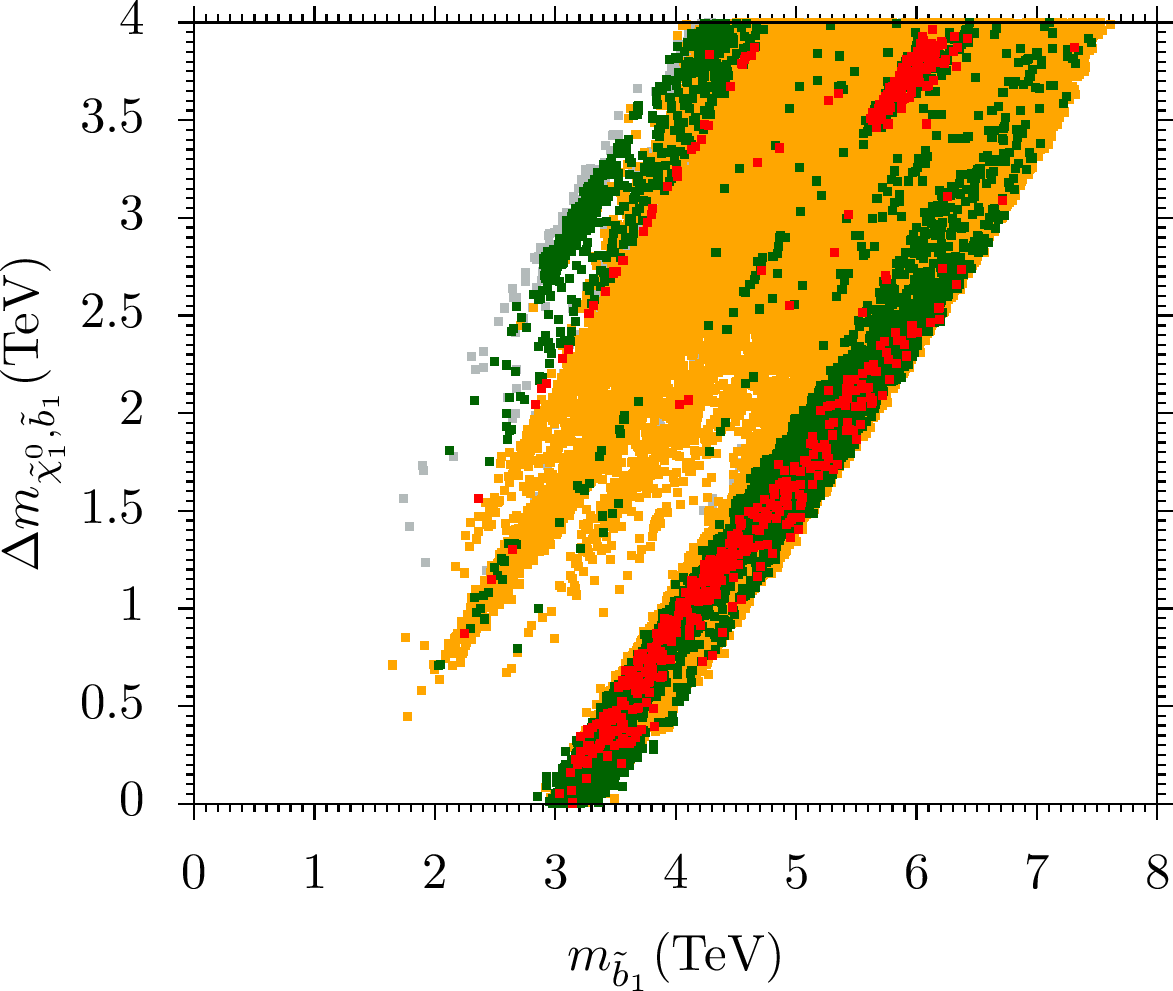}
    \caption{Plots in $m_{\tilde b_{1}}-m_{\tilde \chi_{1}^{0}}$ and $m_{\tilde b_{1}}-\mid \Delta m_{\tilde \chi_{1}^{0},\tilde b_{1}}\mid$ planes,for $\mu <0$ the scenario. Gray points satisfy REWSB with a neutralino LSP. Orange points (subset of gray) obey all LHC SUSY mass limits, Higgs mass constraints, and B-physics bounds, but exhibit relic densities exceeding (over-saturated) the observed value.  Green points (subset of orange) yield relic densities below (under-saturated ) observational limits.  Finally,Red points (subset of green) satisfy the Planck 2018 saturated relic density  (5$\sigma$ bounds). The black line delineates regions where coannihilation mechanisms dominate.
		}
		\label{fig1}
\end{figure*}

Figure~\ref{fig1} illustrates the sbottom-neutralino sector of the parameter space. In the left panel, we display the mass of the next-to-lightest supersymmetric particle (NLSP) sbottom, $m_{\tilde b_1}$, as a function of the lightest neutralino (LSP) mass, $m_{\tilde\chi_1^0}$, while the right panel shows the corresponding mass splitting,
\begin{figure*}[th!]
	\centering \includegraphics[width=8.90cm]{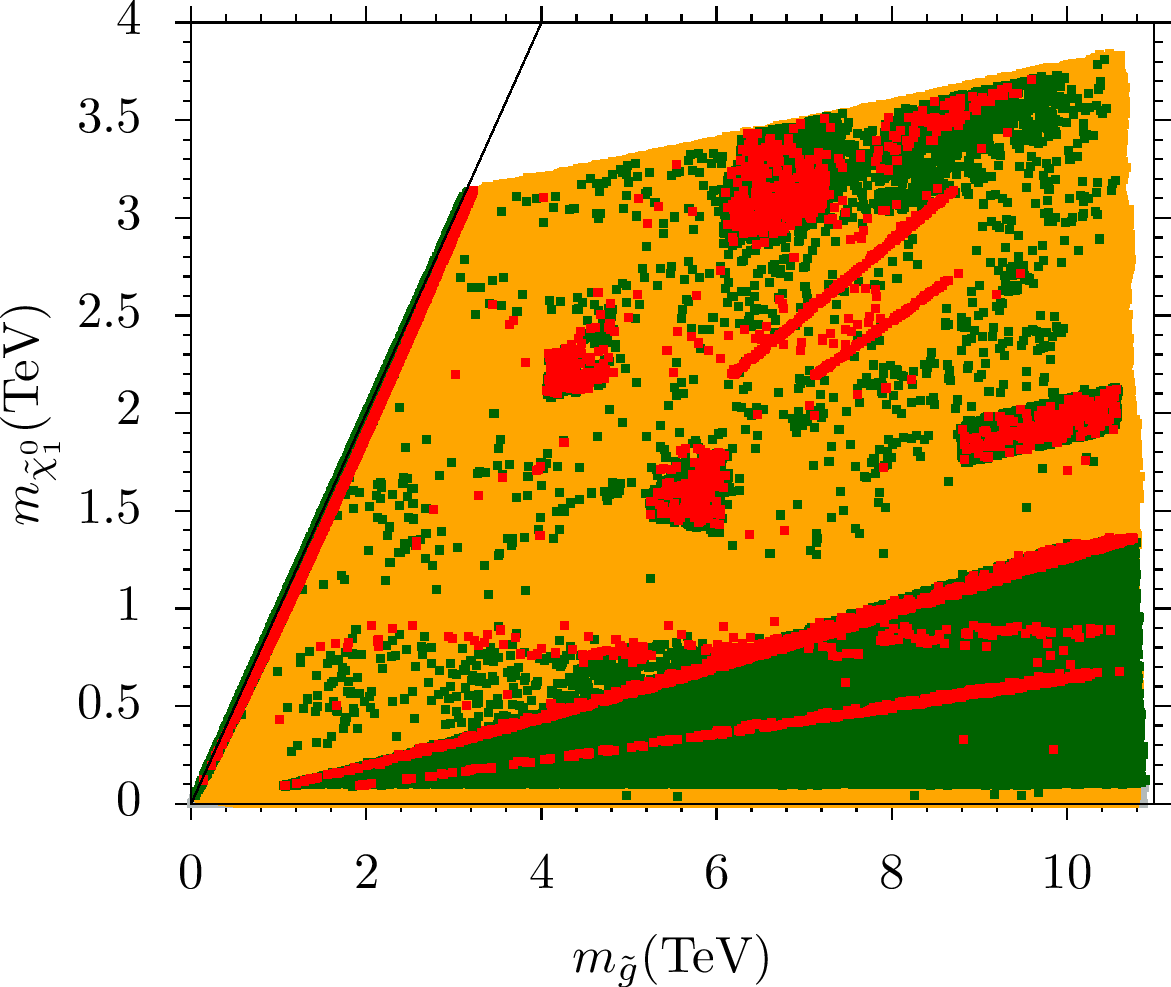}
    \centering \includegraphics[width=8.90cm]{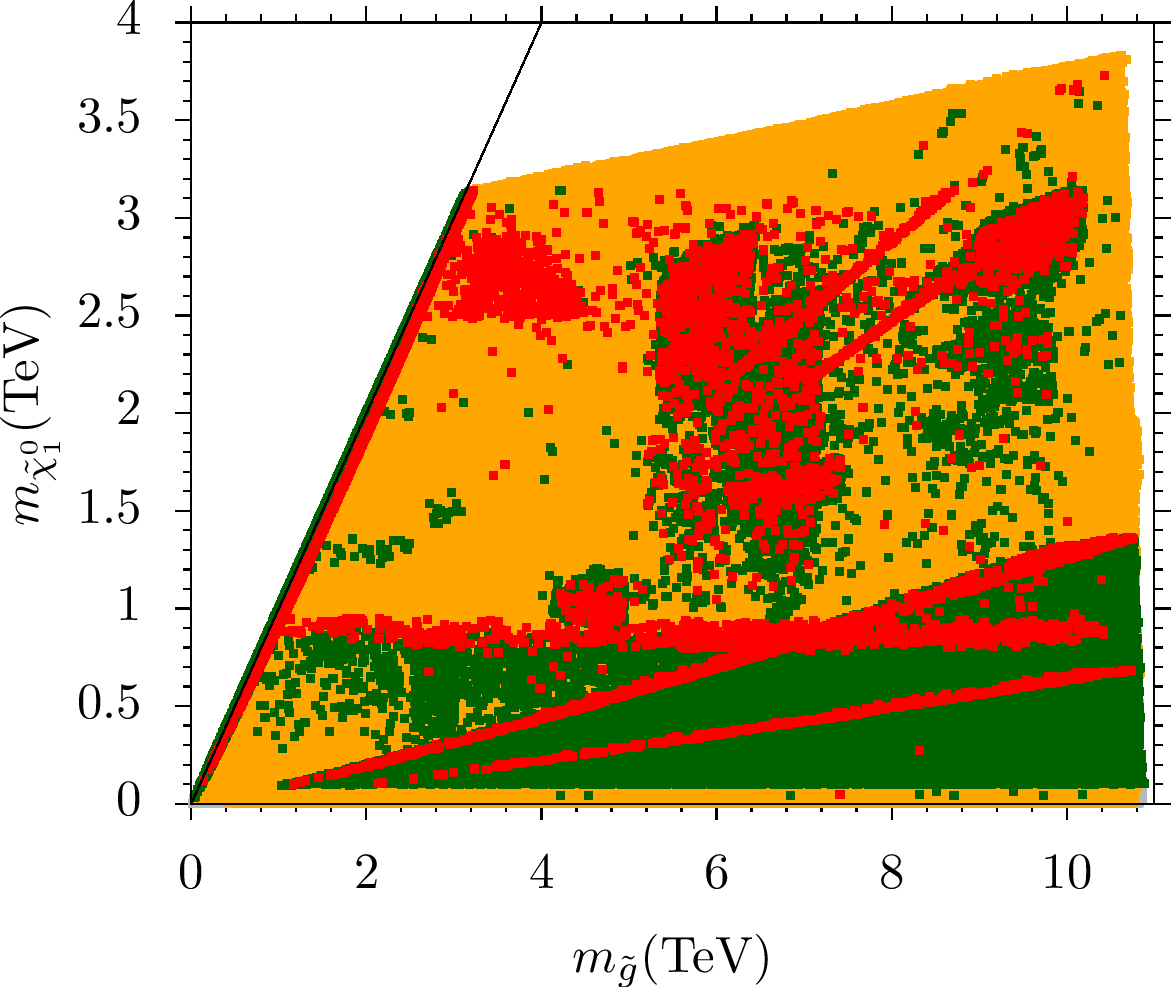}
	\centering \includegraphics[width=8.90cm]{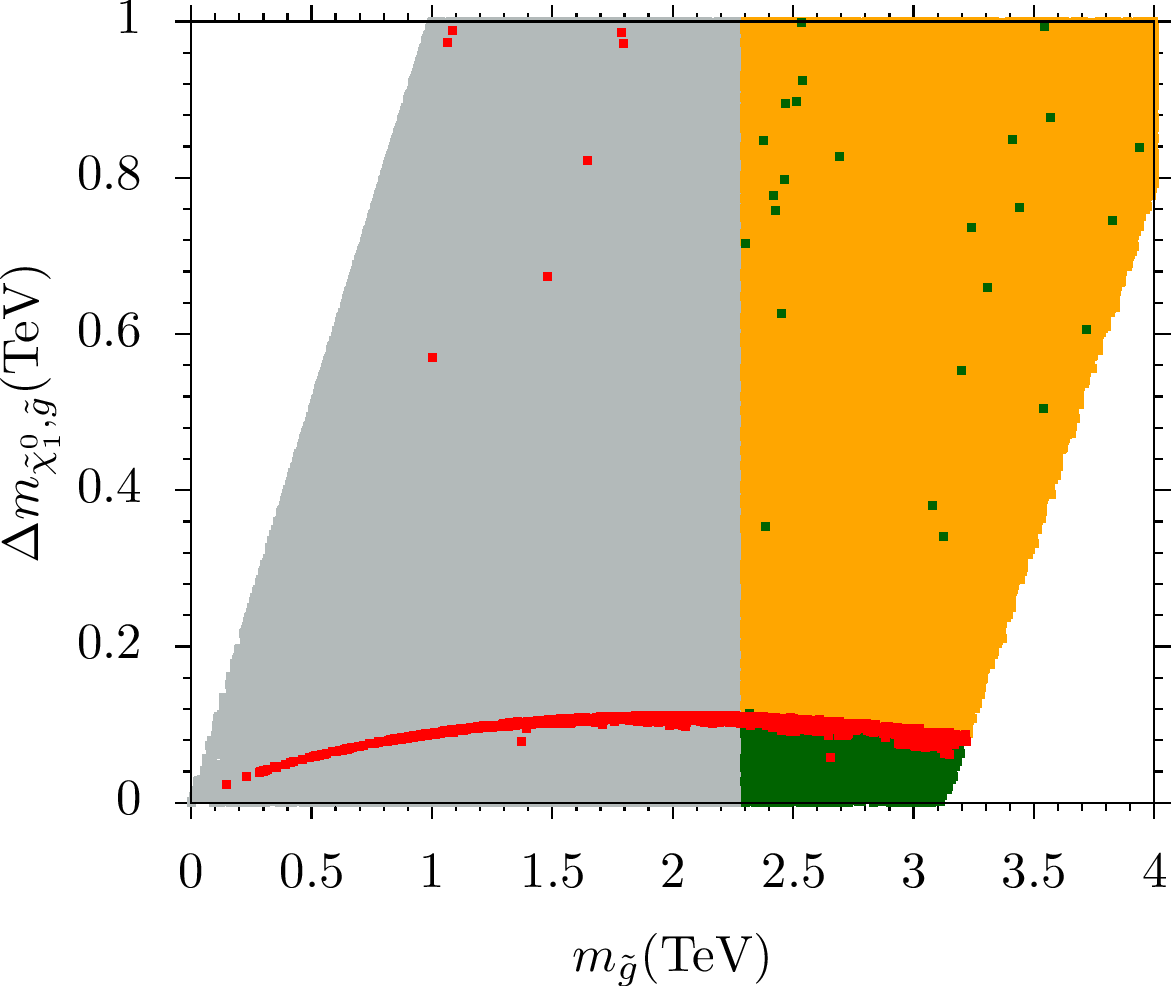}
    \centering \includegraphics[width=8.90cm]{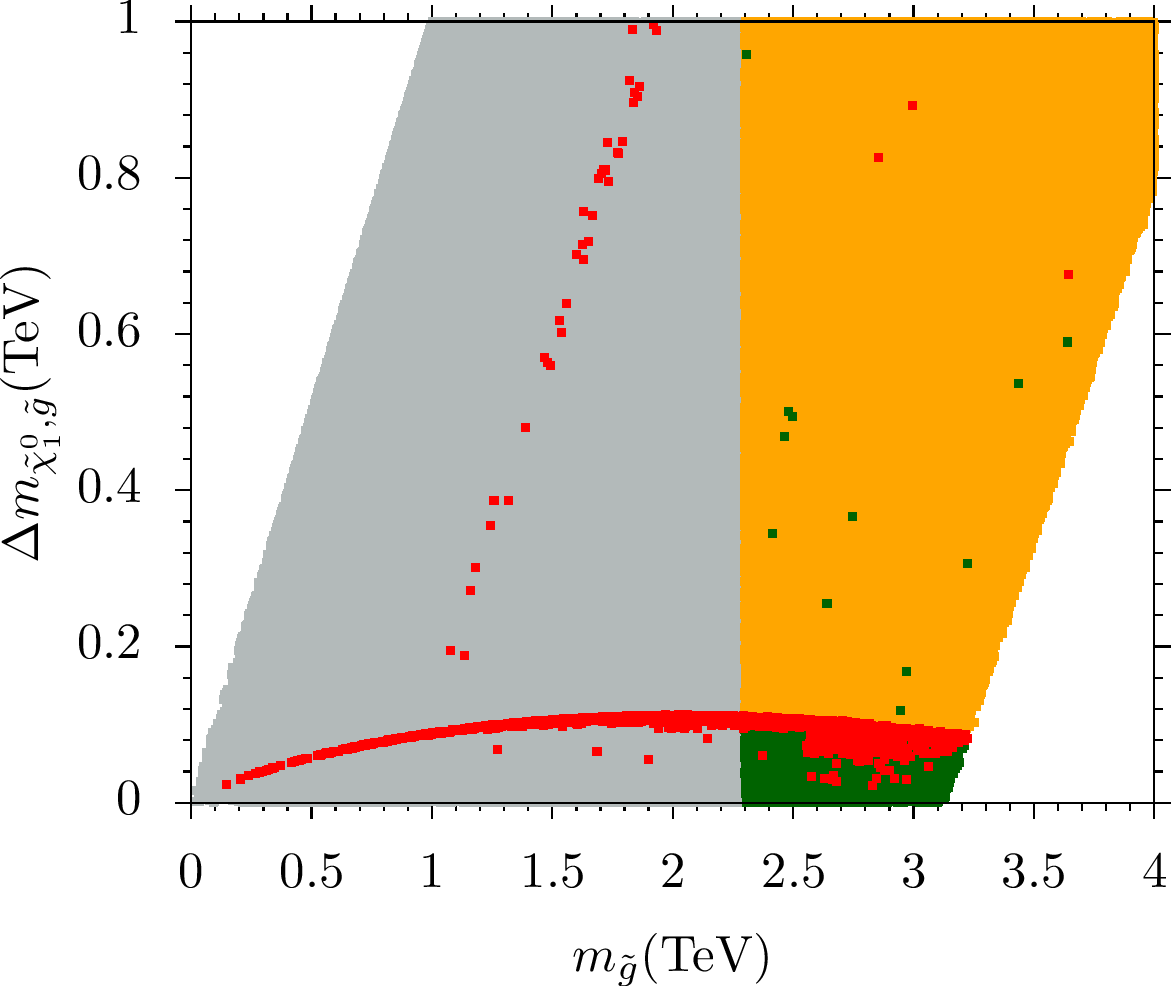}
    \caption{Mass bounds and constraints in the $m_{\tilde g}-m_{\tilde \chi_{1}^{0}}$ and $m_{\tilde g}-\mid \Delta m_{\tilde \chi_{1}^{0},\tilde g}\mid$ planes, for $\mu <0$ (left panels) and $\mu > 0$ (right panels) with the same color scheme as in Fig. \ref{fig1}.
		}
		\label{fig2}
\end{figure*}
$\vert \Delta m_{\tilde b_1,\tilde\chi_1^0} \vert$, plotted against $m_{\tilde\chi_1^0}$.
The color coding is as follows: gray points satisfy REWSB and yield a neutralino LSP; orange points are a subset that additionally satisfy LEP mass limits, Higgs mass, $B$-physics observables, and current LHC sparticle bounds but predict an overabundant relic density; green points satisfy all the above constraints except the relic-density upper bound, resulting in an underabundant dark matter density; and red points satisfy all constraints, including consistency with the Planck-measured relic density.
The diagonal line in the left panel indicates the sbottom-neutralino coannihilation region, where the NLSP sbottom is nearly degenerate in mass with the LSP neutralino. Sbottom coannihilation was first explored in the context of $SU(5)$ grand unification in Ref.~\cite{Gogoladze:2011ug}. In the present work, we demonstrate for the first time that sbottom-neutralino coannihilation solutions consistent with the Planck 2018 $5\sigma$ relic density bound and all current experimental constraints can be realized within the supersymmetric Pati-Salam model without considering the Yukawa unification. \footnote{As discussed earlier, NLSP sbottom solutions are absent for $\mu>0$ in our scans.} To the best of our knowledge, such solutions have previously been reported only in the context of generalized minimal supergravity models~\cite{Khan:2025yit} and, more recently, in bottom-$\tau$ Yukawa coupling unification~\cite{Ahmed:2022ibc}.
From the left panel of Fig.~\ref{fig1}, we observe that the sbottom NLSP solutions consistent with the dark matter relic density (red points) lie in the mass range
$2.7~\text{TeV} \lesssim m_{\tilde b_1} \lesssim 3.4~\text{TeV}$.
The right panel displays the mass difference $\Delta m_{\tilde b_1,\tilde\chi_1^0}$ as a function of the NLSP sbottom mass. Throughout this analysis, we impose the requirement
$
\frac{\Delta m_{\text{NLSP,LSP}}}{m_{\text{LSP}}} \lesssim 10\%,
\quad
\Delta m_{\text{NLSP,LSP}} \equiv m_{\text{NLSP}} - m_{\text{LSP}},$
ensuring efficient coannihilation. Accordingly, the red points with small mass splittings correspond to viable sbottom NLSP solutions. We comment here that when the mass difference exceeds the bottom-quark mass, the dominant collider signature arises from direct sbottom pair production,
$pp \rightarrow \tilde b_1 \tilde b_1^{\ast} + X 
\rightarrow b\bar b + \cancel{E}_{inv},
$
where $\tilde b_1 \rightarrow b \tilde\chi_1^0$. Furthermore, the same sign sbottom pair production processes, $\tilde b_1 \tilde b_1$ and $\tilde b_1^{\ast} \tilde b_1^{\ast}$, may contribute.
Current LHC searches already place important constraints on light sbottom scenarios. For example, ATLAS has excluded sbottom masses up to $\sim 1.5~\text{TeV}$ and $\sim 0.85~\text{TeV}$ in decay chains involving $\tilde b_1 \rightarrow b \tilde\chi_2^0 \rightarrow b h \tilde\chi_1^0$ for $\Delta m_{\tilde\chi_2^0,\tilde\chi_1^0} = 130~\text{GeV}$~\cite{ATLAS:2019gdh,ATLAS:2021pzz}. Similarly, sbottom masses up to $\sim 1.6~\text{TeV}$ maybe excluded for $\tilde b_1 \rightarrow t \tilde\chi_2^{0}$ with $\Delta m_{\tilde\chi_1^{\pm},\tilde\chi_1^0} = 100~\text{GeV}$~\cite{ATLAS:2019fag}. In the simplified topology $\tilde b_1 \rightarrow b \tilde\chi_1^0 (b-jets+\cancel{E})$, NLSP sbottom masses up to $\sim 1.27~\text{TeV}$ for massless neutralinos can be excluded. 
In the compressed regime, $m_{\tilde b_1} \simeq m_{\tilde\chi_1^0}$, dedicated searches exploiting displaced vertices and soft $b$-jet signatures exclude NLSP sbottom masses up to $\sim 660~\text{GeV}$ for $\Delta m_{\tilde b_1,\tilde\chi_1^0} \sim 10~\text{GeV}$~\cite{ATLAS:2021yij}, while monojet-based analyses constrain sbottom masses up to $\sim 600~\text{GeV}$. Notably, for $m_{\tilde b_1} \gtrsim 800~\text{GeV}$, current analyses do not impose significant limits~\cite{ATLAS:2021yij}.
As we can see, in the first two cases, sbottom is not the NLSP, but the last two channels
are relevant. So, our results are safe. Therefore, our solution will be accessible to LHC Run-3 and future colliders.
\begin{figure*}[th!]
	\centering \includegraphics[width=8.90cm]{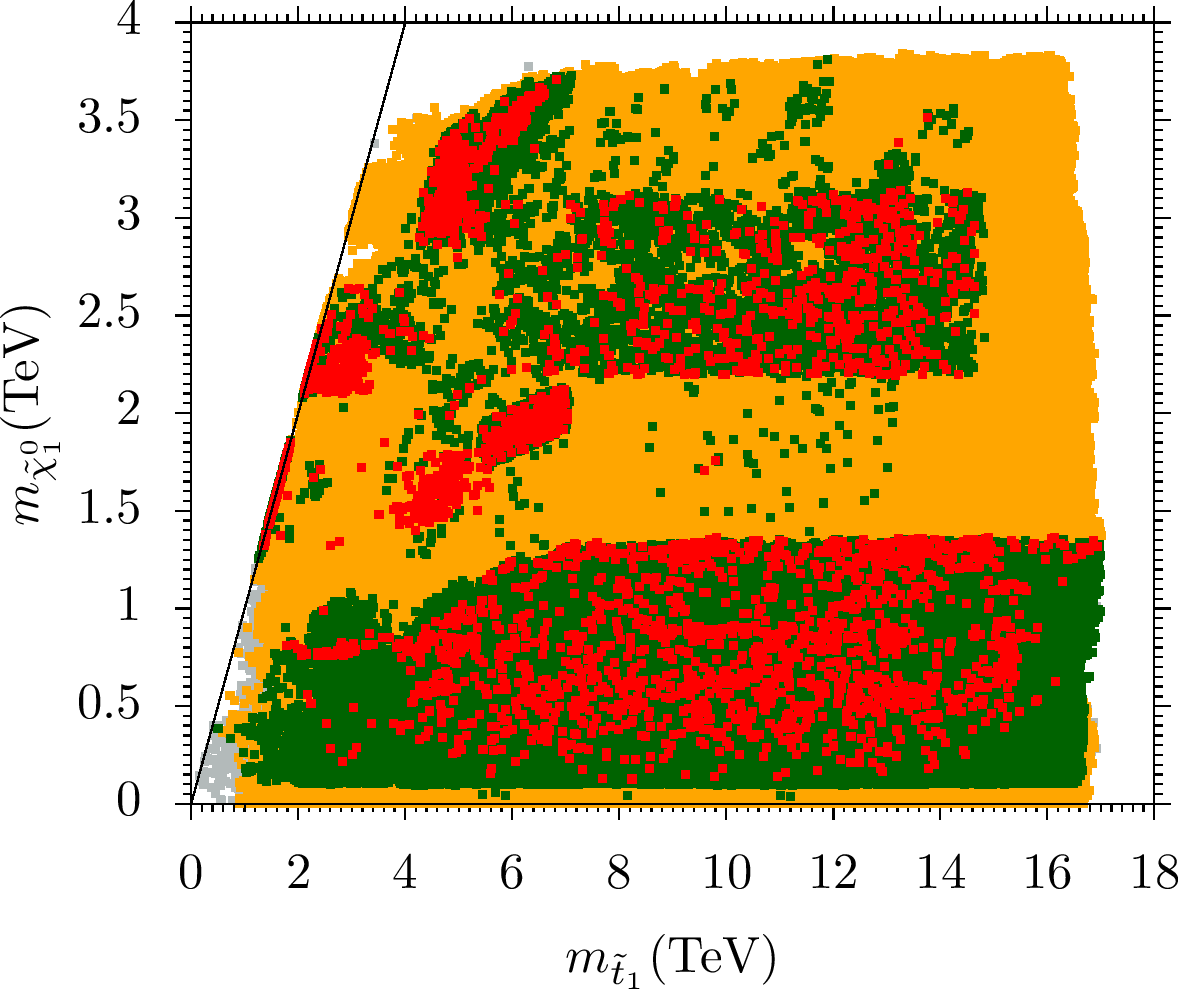}
    \centering \includegraphics[width=8.90cm]{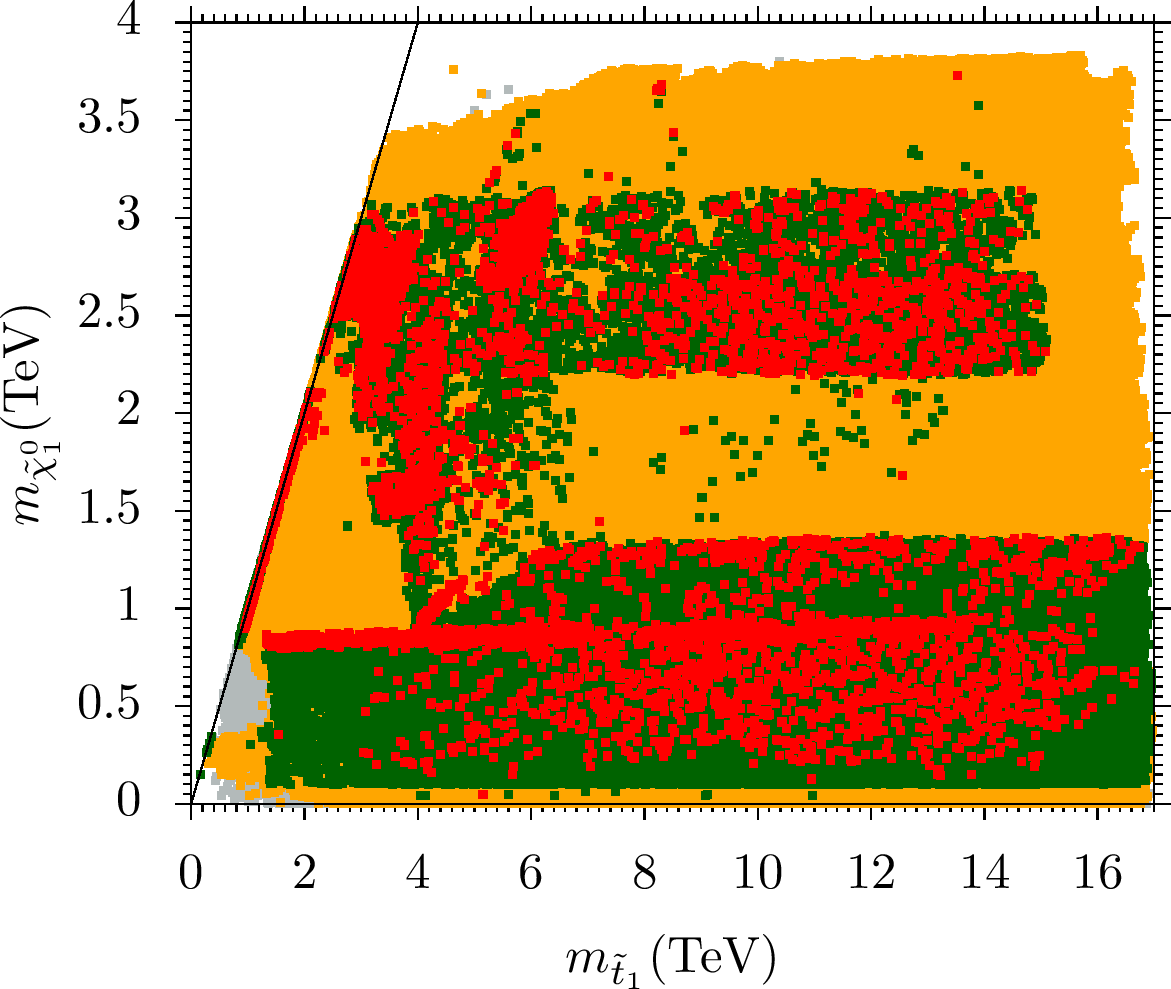}
	\centering \includegraphics[width=8.90cm]{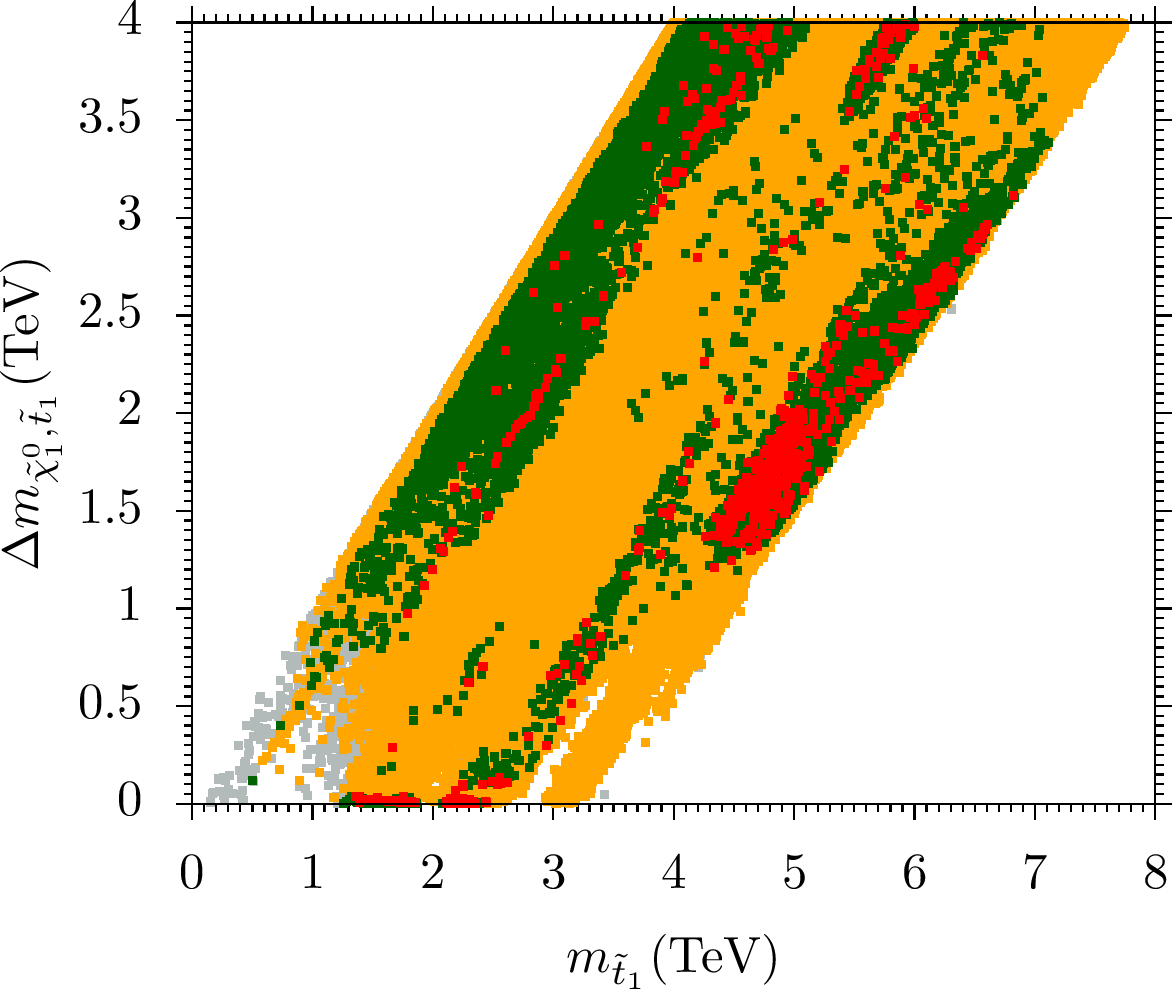}
    \centering \includegraphics[width=8.90cm]{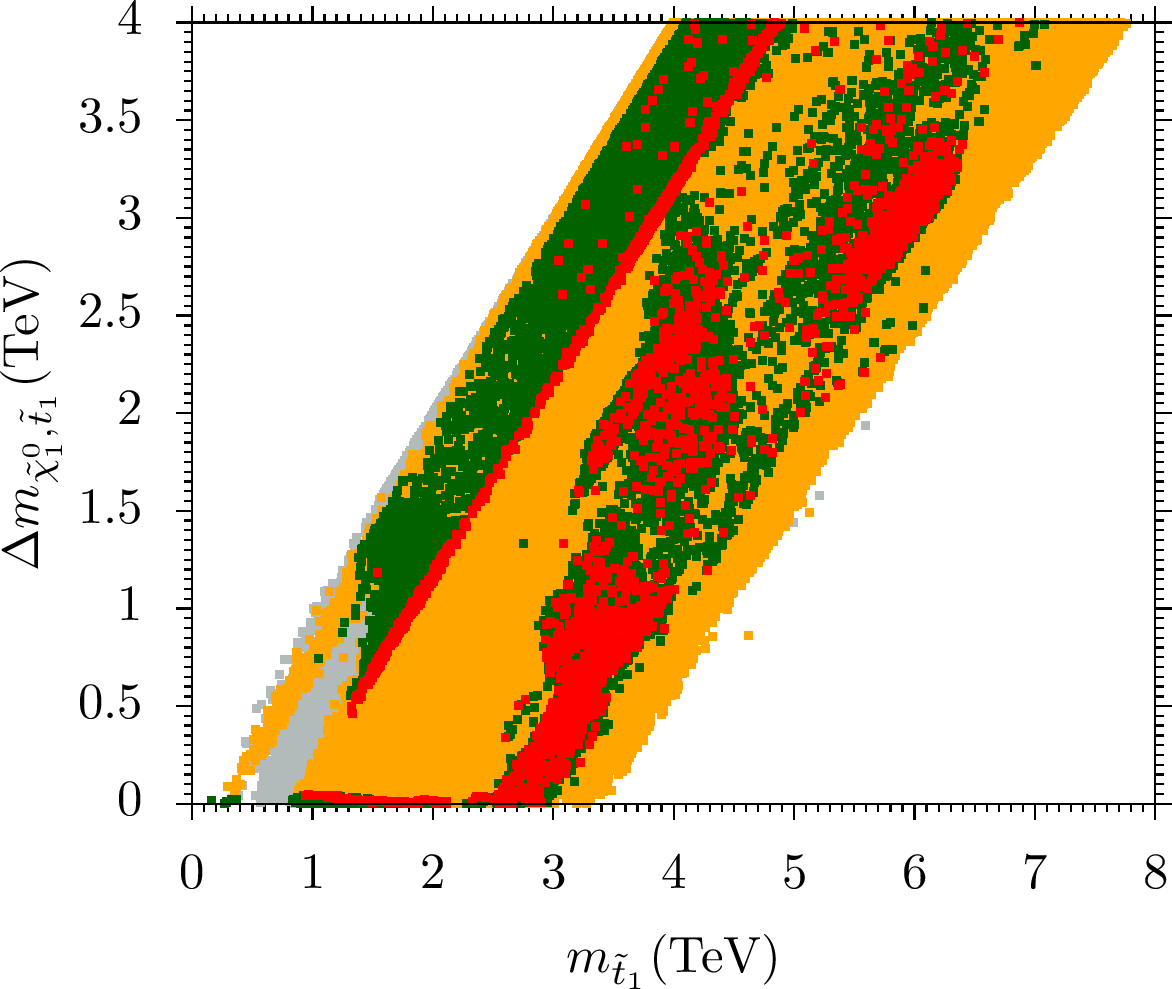}
    \caption{Mass bounds and constraints in the $m_{\tilde t_1}$-$m_{\tilde \chi_{1}^{0}}$ and $m_{\tilde t_1}$-$\vert \Delta m_{\tilde \chi_{1}^{0},\tilde t_{1}}\vert$ planes, for $\mu <0$ (left panels) and $\mu > 0$ (right panels) with the same color scheme as in Fig. \ref{fig1}.
		}
		\label{fig3}
\end{figure*}

Figure~\ref{fig2} presents the gluino-neutralino coannihilation region. The upper panels show the lightest neutralino mass, $m_{\tilde{\chi}_1^0}$, versus the gluino NLSP mass, $m_{\tilde g}$. The left (right) panel corresponds to the $\mu<0$ ($\mu>0$) scenario. The bottom panels show the associated mass splitting,
$\lvert \Delta m_{\tilde g,\tilde{\chi}_1^0} \rvert \equiv \lvert m_{\tilde g} - m_{\tilde{\chi}_1^0} \rvert$,
plotted against $m_{\tilde g}$. The color coding follows that of Fig.~\ref{fig1}; except for the gluino mass bounds discussed in Sec.~\ref{sec:scan}, which are not imposed here in order to highlight the full kinematic structure of the coannihilation region.
The red points aligned along the diagonal in the top panel correspond to scenarios in which the neutralino and gluino are nearly degenerate in mass and belong to gluino-neutralino coannihilation. Notably, the NLSP gluino mass in this region extends from approximately $0.1~\text{TeV}$ up to $3.2~\text{TeV}$. This represents a substantial extension of the viable parameter space compared to earlier studies, where the gluino mass was typically limited to around $1~\text{TeV}$~\cite{Raza:2014upa}, and also exceeds the maximum value of $2.6~\text{TeV}$ reported in Ref.~\cite{Gomez:2020gav}. 
The down panel further clarifies the compressed nature of the spectrum: the diagonal red points exhibit a gluino-neutralino mass splitting $\Delta m_{\tilde{g},\tilde{\chi}_1^0} \lesssim 100~\text{GeV}$, while the remaining red points correspond to configurations that depart from the coannihilation strip. In this highly compressed spectrum, the dominant gluino decay channel is $\tilde{g} \to b\bar{b} \tilde{\chi}_1^0$, one that proceeds via off-shell third-generation squarks. Although alternative decay modes are kinematically allowed, they are phenomenologically less relevant due to severe background contamination, particularly in final states involving light quarks with low transverse momentum.
The compressed spectrum implies that jets originating from light quarks often fail to produce sufficiently energetic tracks, rendering such signals difficult to distinguish from Standard Model backgrounds. In contrast, final states containing $b$ quarks benefit from the presence of secondary vertices and displaced tracks, which significantly enhance reconstruction efficiency and background rejection. As a result, searches targeting $b$ track jets are especially well suited for probing this scenario. Current ATLAS analyses of compressed gluino-neutralino spectra constrain the gluino mass to be above approximately $1.2~\text{TeV}$ in the nearly degenerate limit~\cite{ATLAS:2018yhd,ATLAS:2021ilc}. While a portion of the parameter space identified here is therefore already excluded, a substantial region remains viable and consistent with existing experimental bounds.
These findings demonstrate that the gluino-neutralino coannihilation scenario in the SUSY Pati-Salam model remains a compelling target for LHC Run-3 and future colliders.

\begin{figure*}[th!]
	\centering \includegraphics[width=8.90cm]{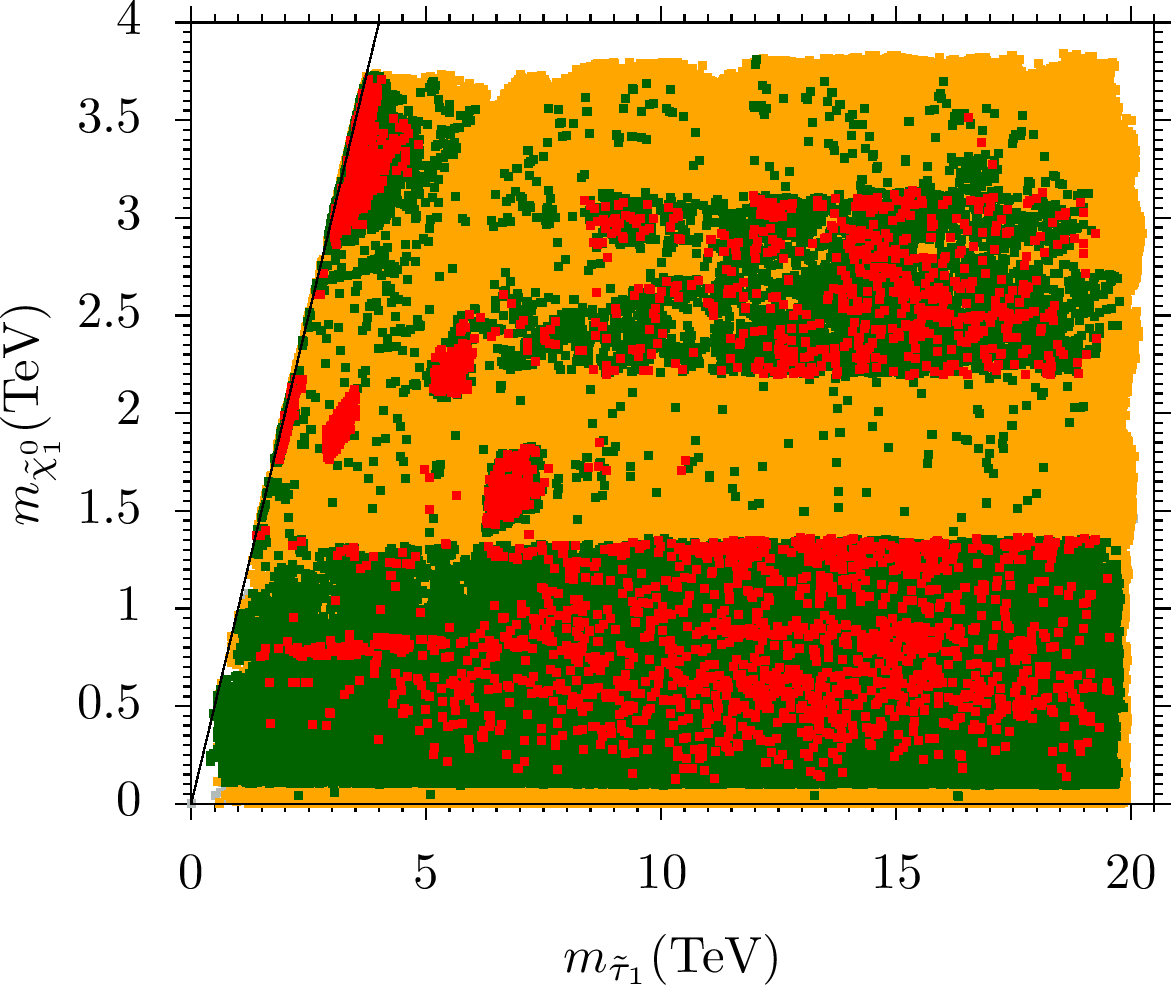}
    \centering \includegraphics[width=8.90cm]{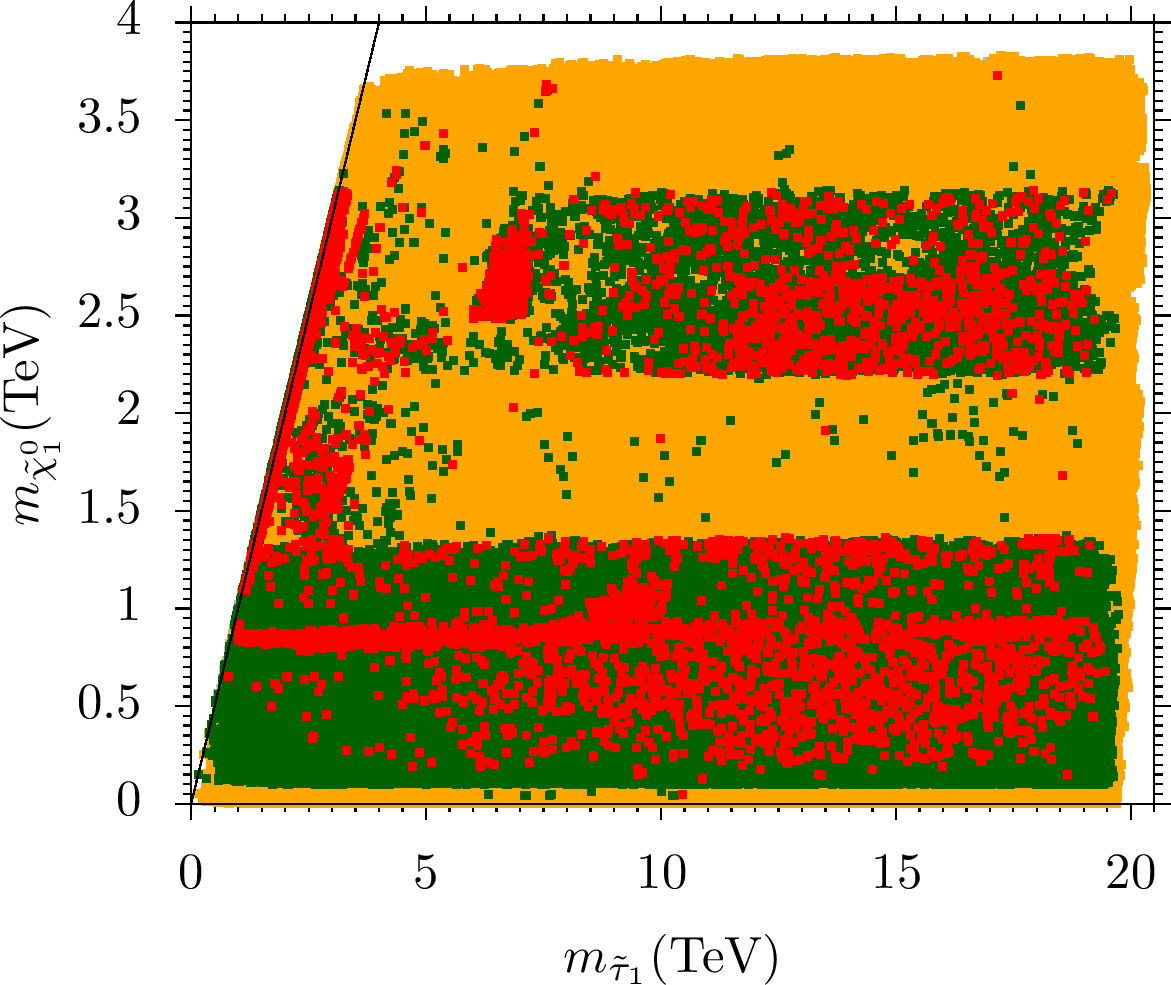}
	\centering \includegraphics[width=8.90cm]{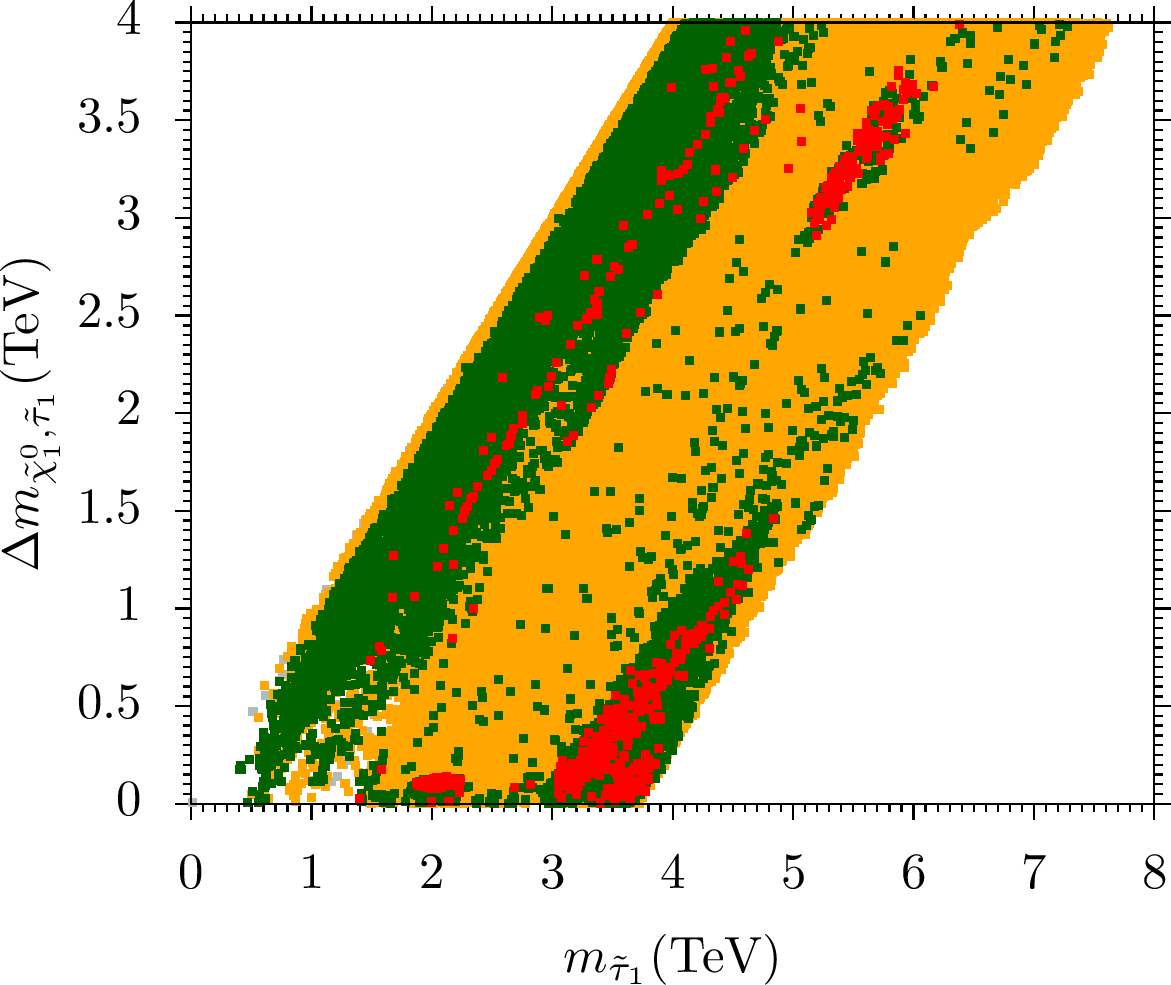}
    \centering \includegraphics[width=8.90cm]{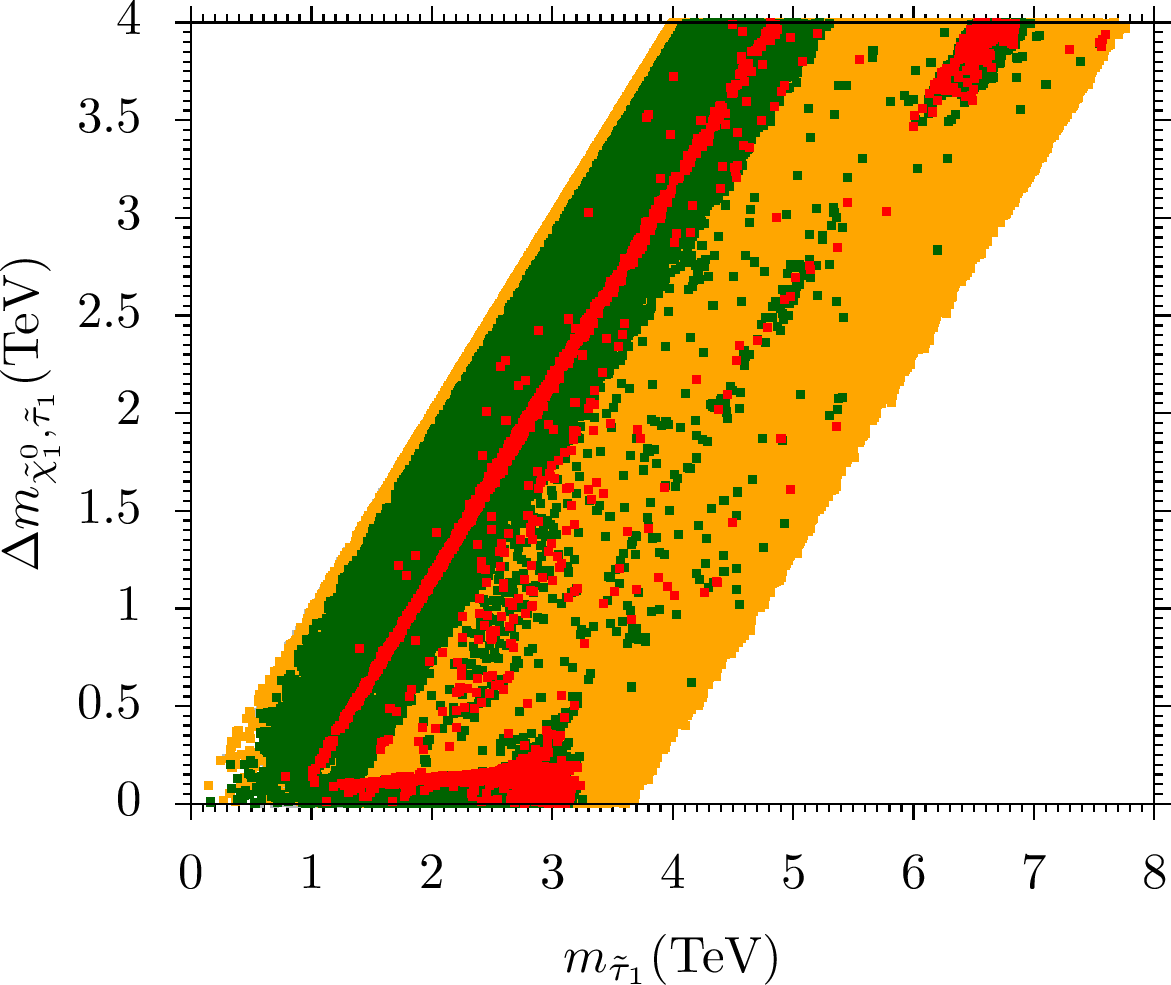}
    \caption{Mass bounds and constraints in the $m_{\tilde \tau_1}$-$m_{\tilde \chi_{1}^{0}}$ and $m_{\tilde \tau_{1}}$-$\vert \Delta m_{\tilde \chi_{1}^{0},\tilde \tau_{1}}\vert$ planes, for $\mu <0$ (left panels) and $\mu > 0$ (right panels) with the same color scheme as in Fig. \ref{fig1}.
		}
		\label{fig4}
\end{figure*}

\begin{figure*}[th!]
	\centering \includegraphics[width=8.90cm]{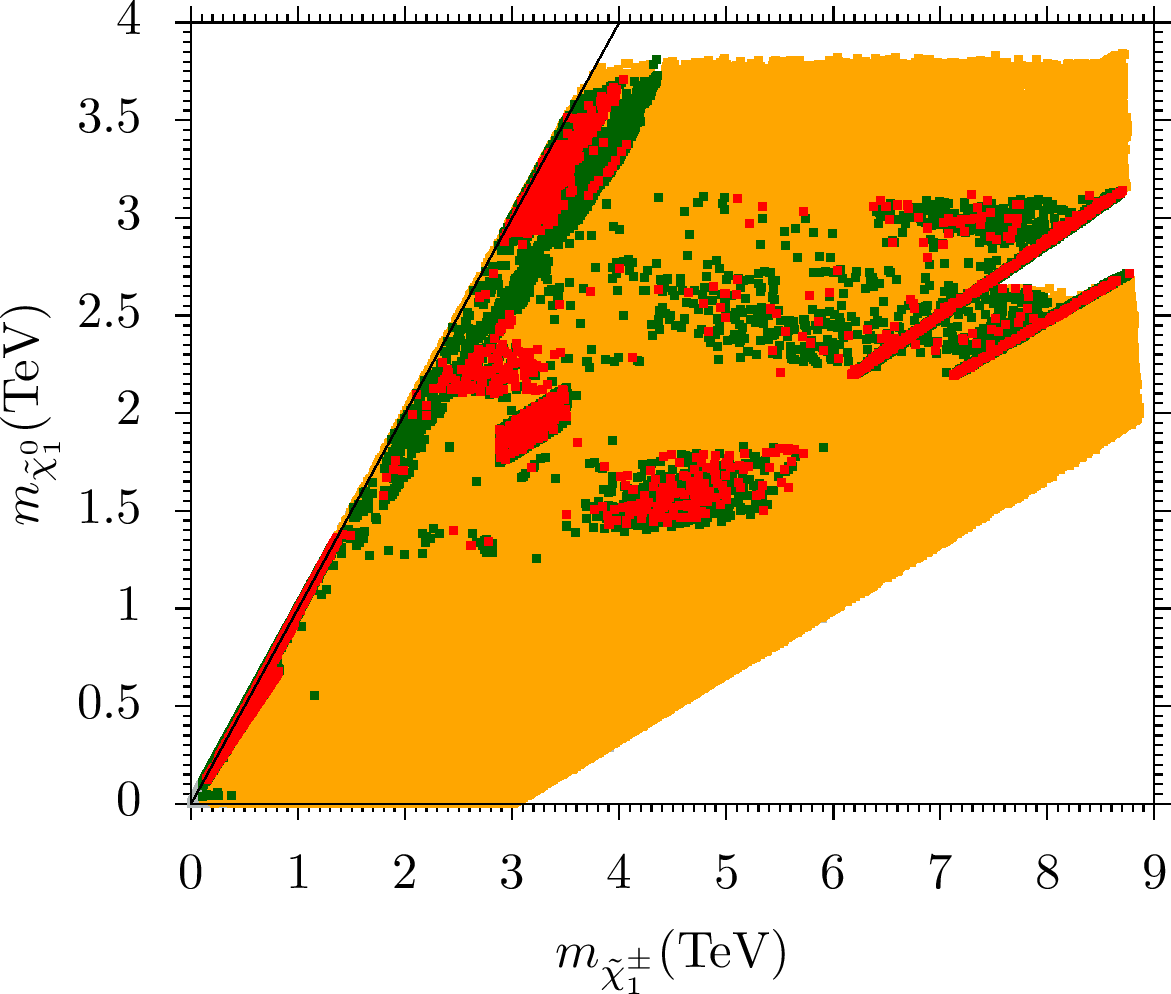}
    \centering \includegraphics[width=8.90cm]{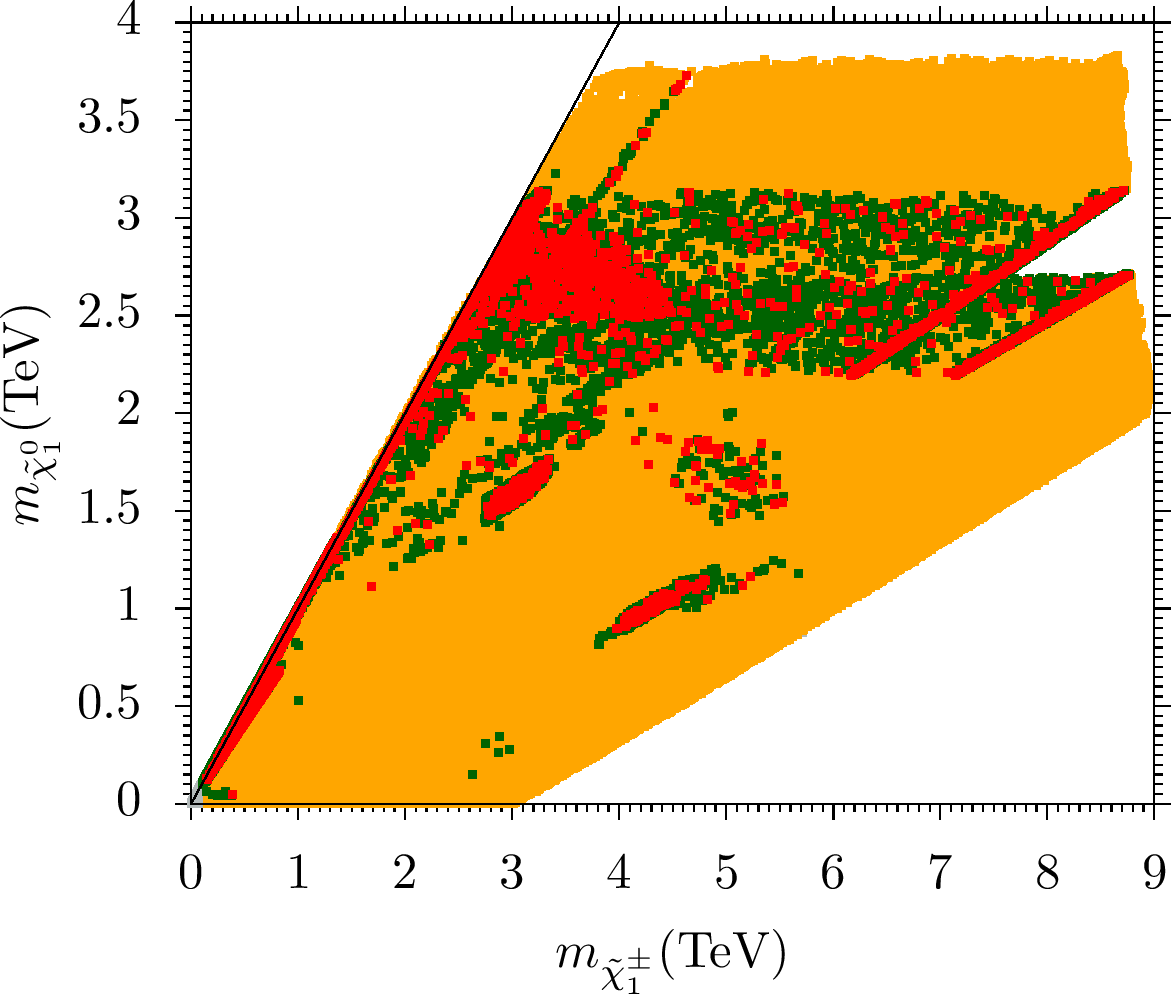}
	\centering \includegraphics[width=8.90cm]{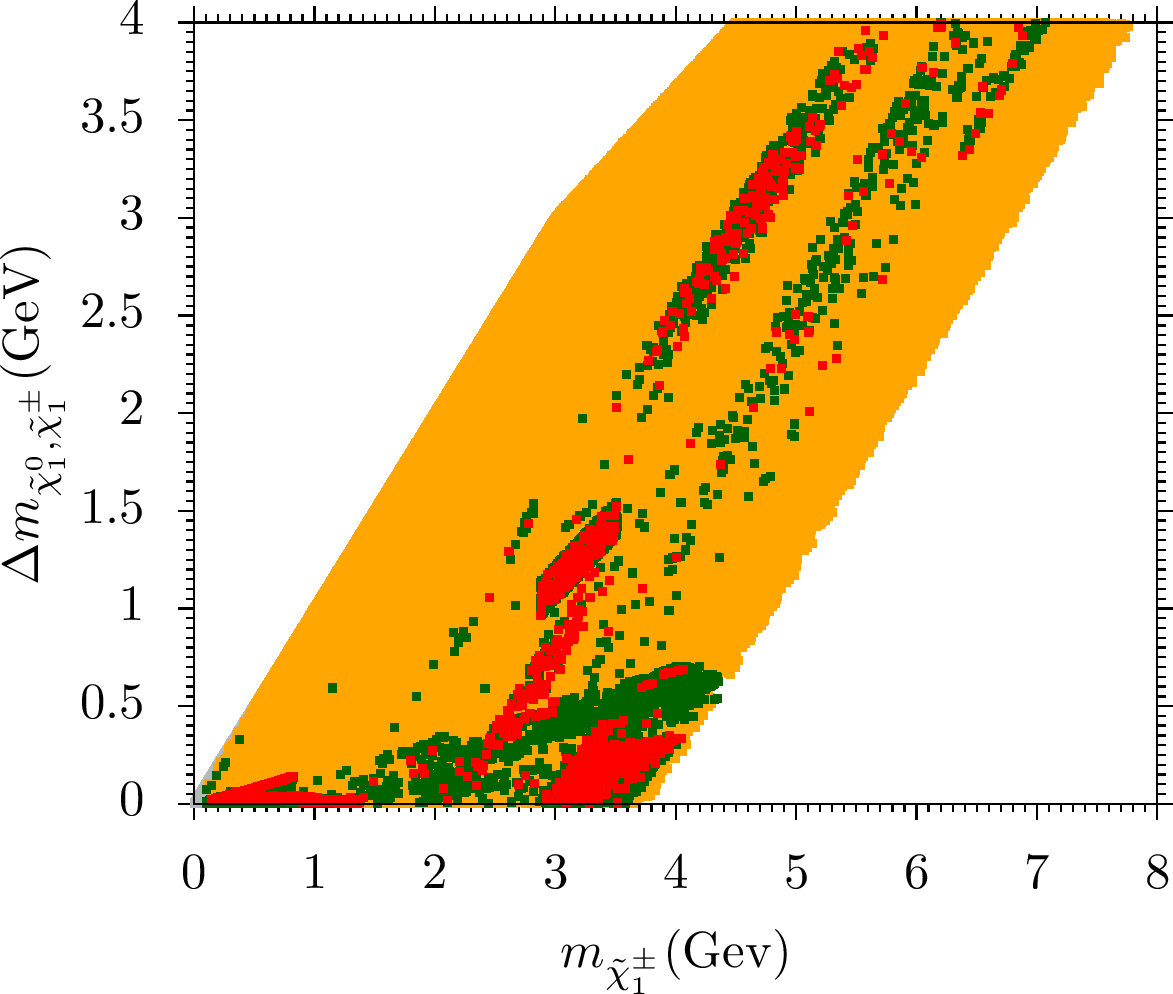}
    \centering \includegraphics[width=8.90cm]{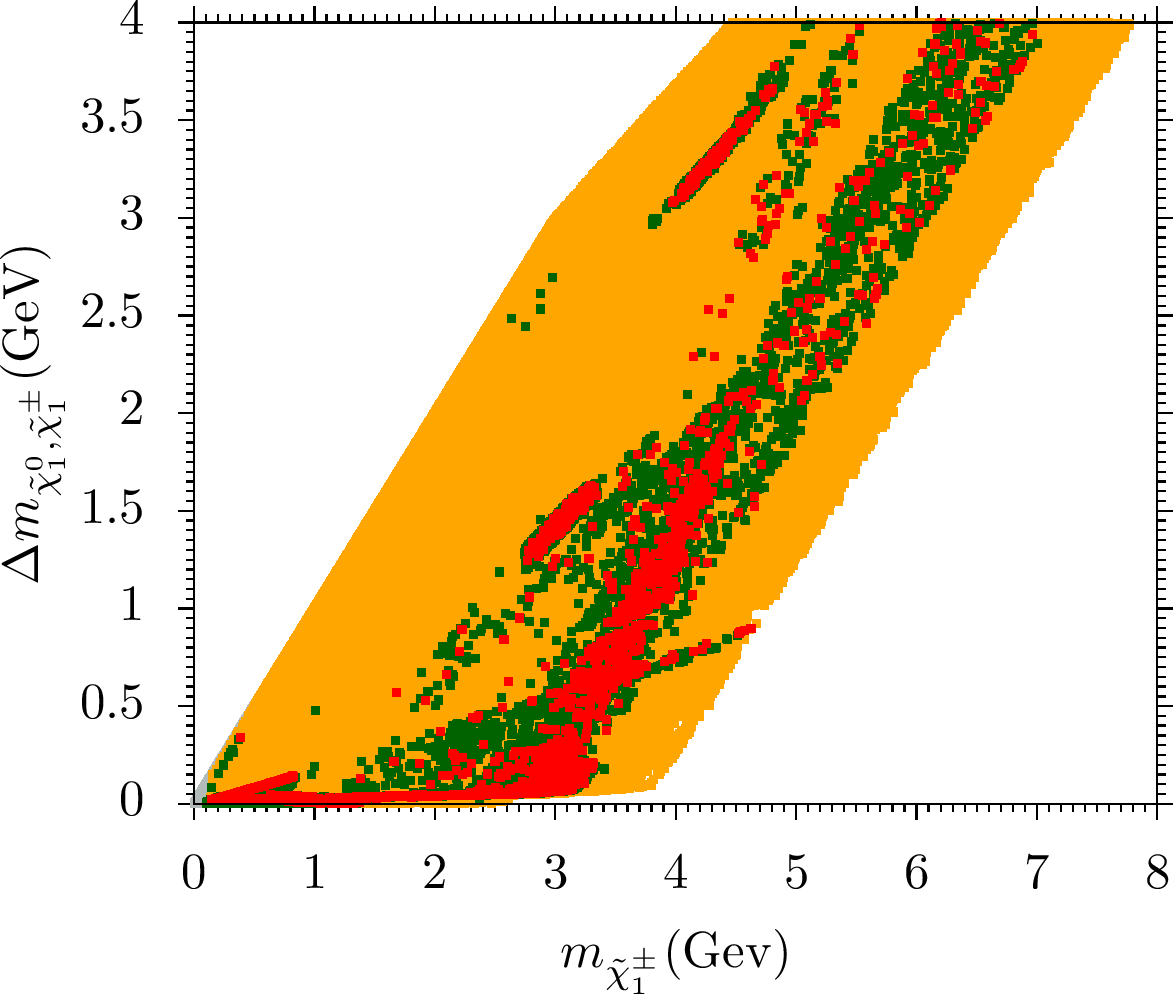}
    \caption{Plots in the $m_{\tilde \chi_{1}^{\pm}}-m_{\tilde \chi_{0}^{\pm}}$ and $m_{\tilde \chi_{1}^{\pm}}-\mid \Delta m_{\tilde \chi_{1}^{0},\tilde \chi_{1}^{\pm}}\mid$ planes, for $\mu <0$ (left panels) and $\mu > 0$ (right panels). The color coding is the same as in Fig. \ref{fig1}.
		}
		\label{fig5}
\end{figure*}

\begin{figure*}[th!]
	\centering \includegraphics[width=8.90cm]{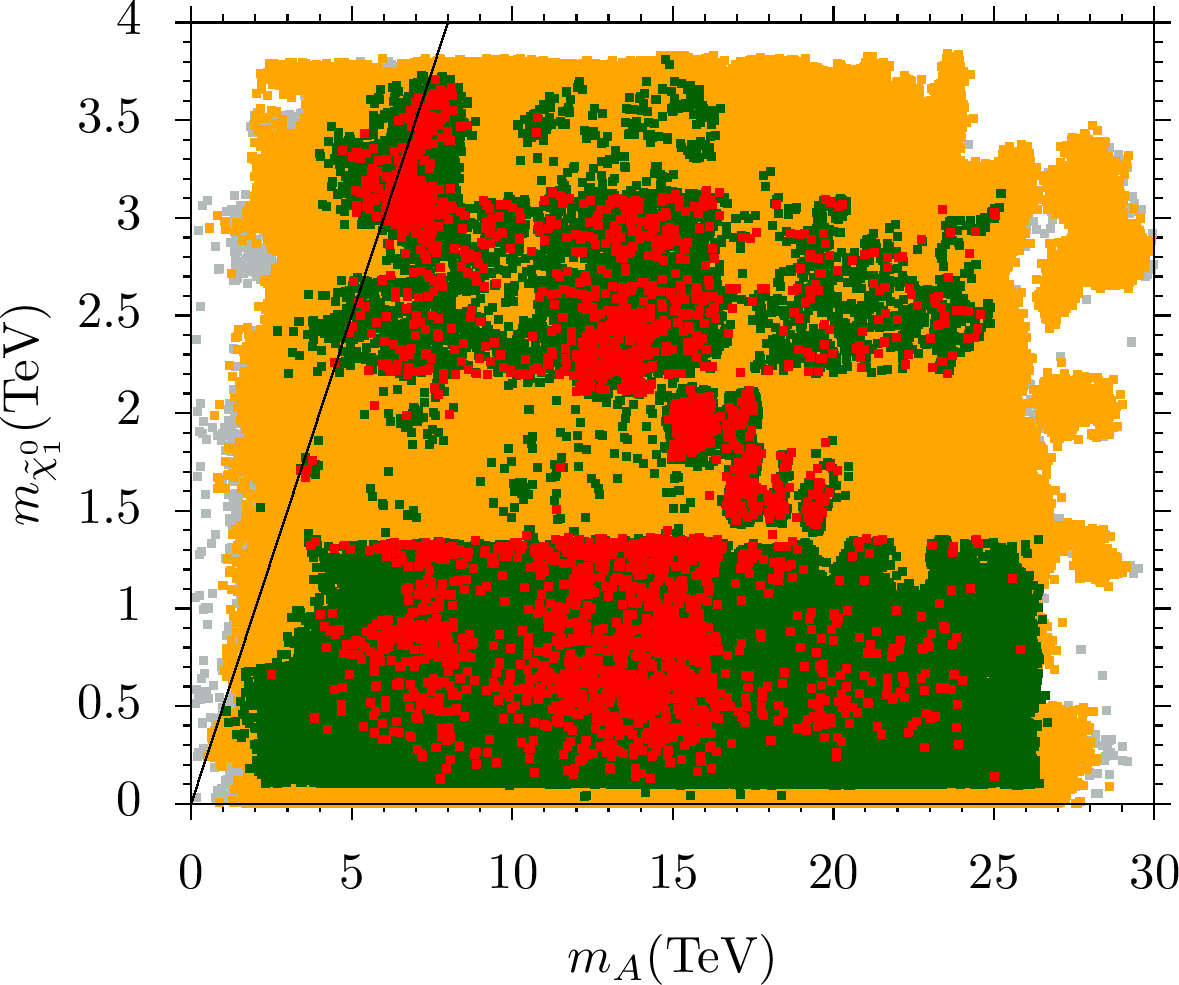}
    \centering \includegraphics[width=8.90cm]{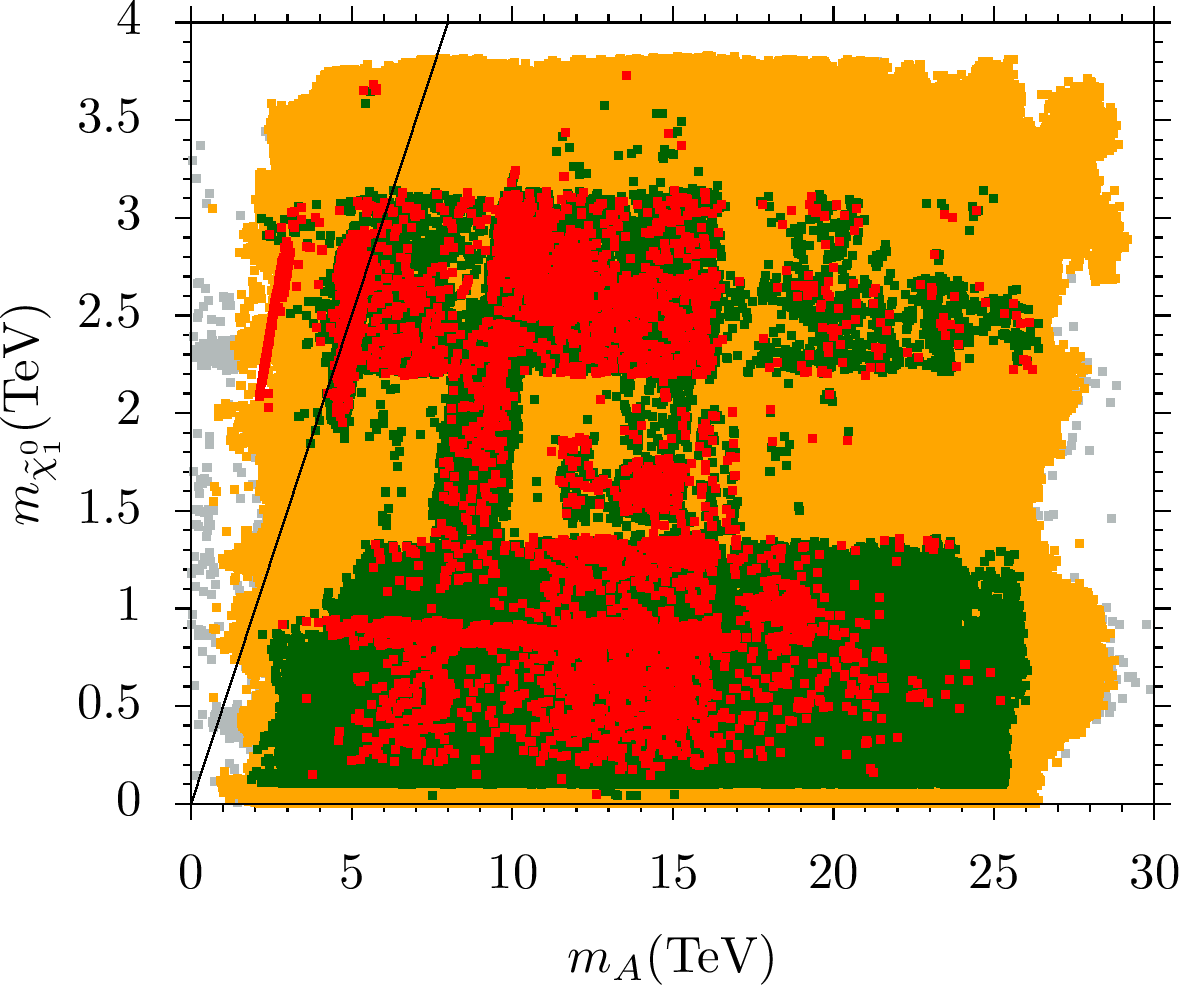}
	\centering \includegraphics[width=8.90cm]{MAn.png}
    \centering \includegraphics[width=8.90cm]{MAp.png}
    \caption{Plots in the $m_{A}-m_{\tilde \chi_{1}^{0}}$ and $m_{A}-\tan\beta$ planes, for $\mu <0$ (left panels) and $\mu > 0$ (right panels). The color coding is the same as in Fig. \ref{fig1}.
		}
		\label{fig6}
\end{figure*}

\begin{figure*}[th!]
	\centering \includegraphics[width=8.90cm]{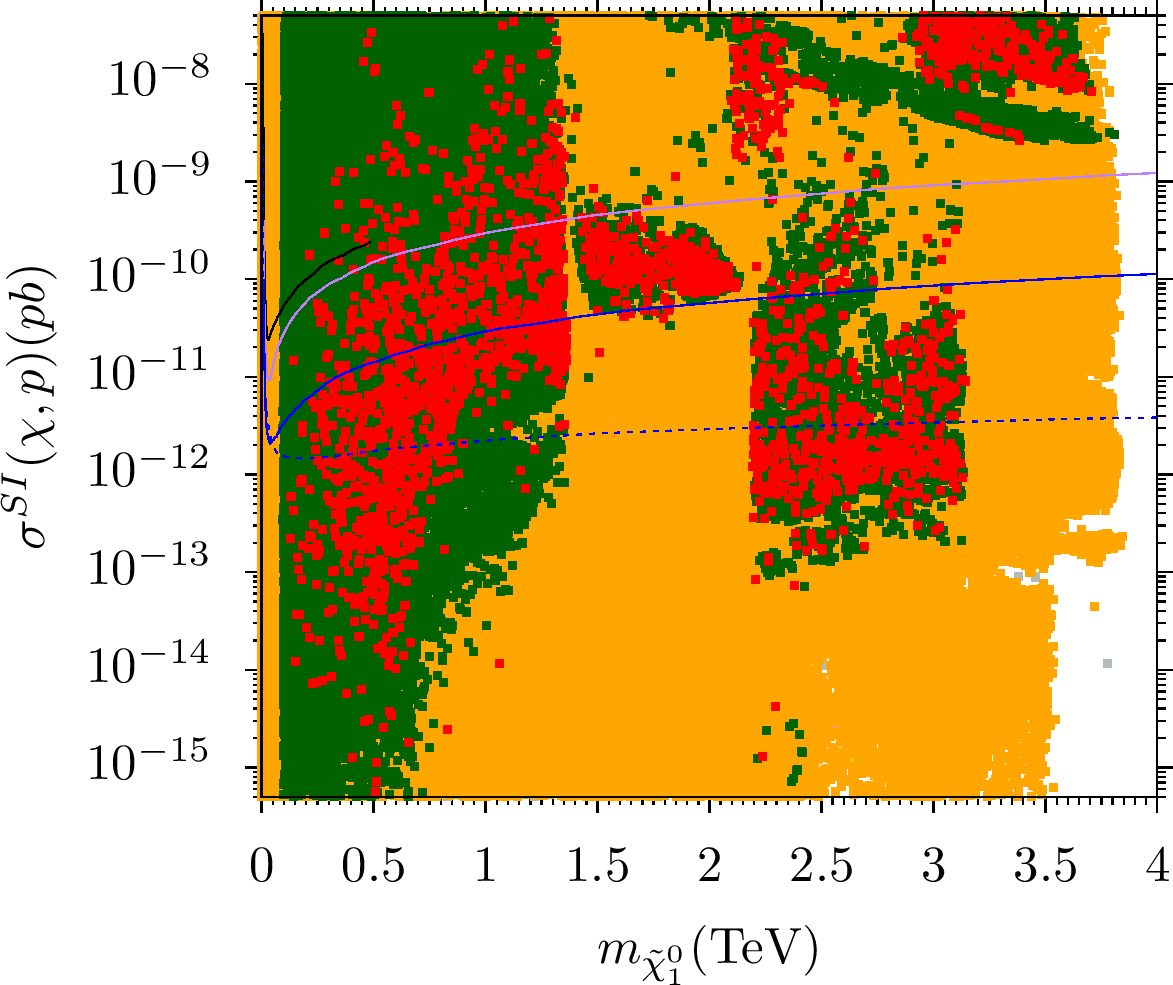}
    \centering \includegraphics[width=8.90cm]{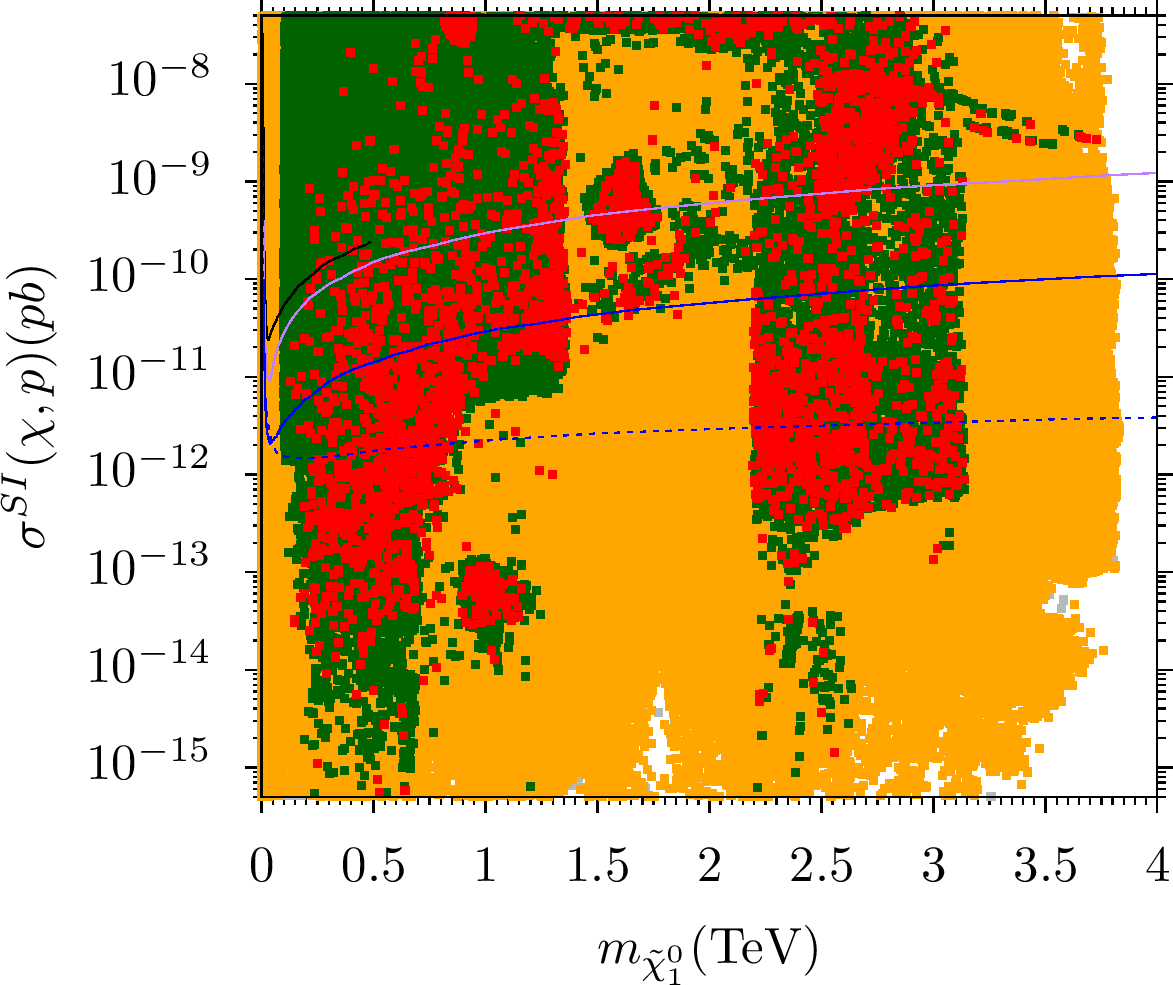}
	\centering \includegraphics[width=8.90cm]{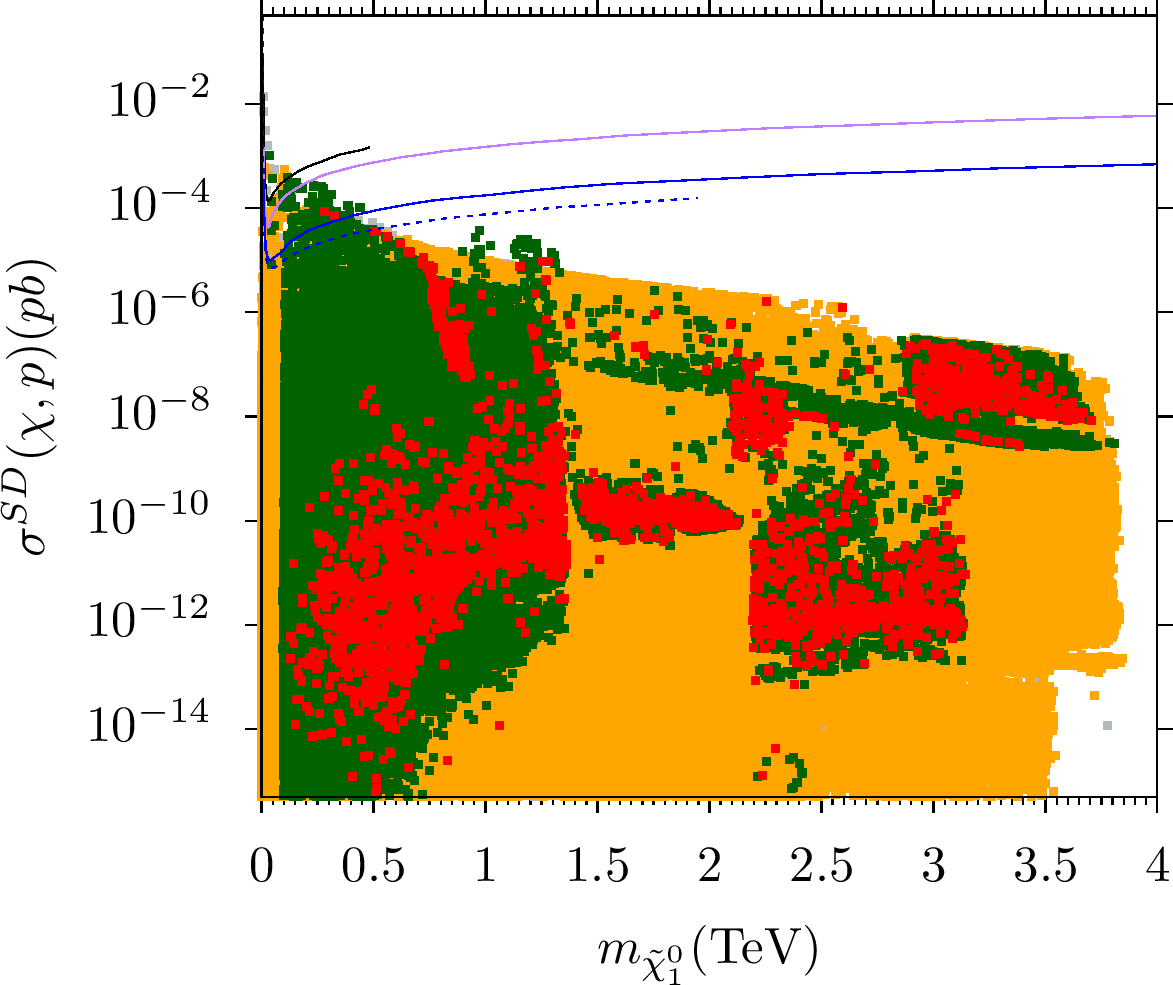}
    \centering \includegraphics[width=8.90cm]{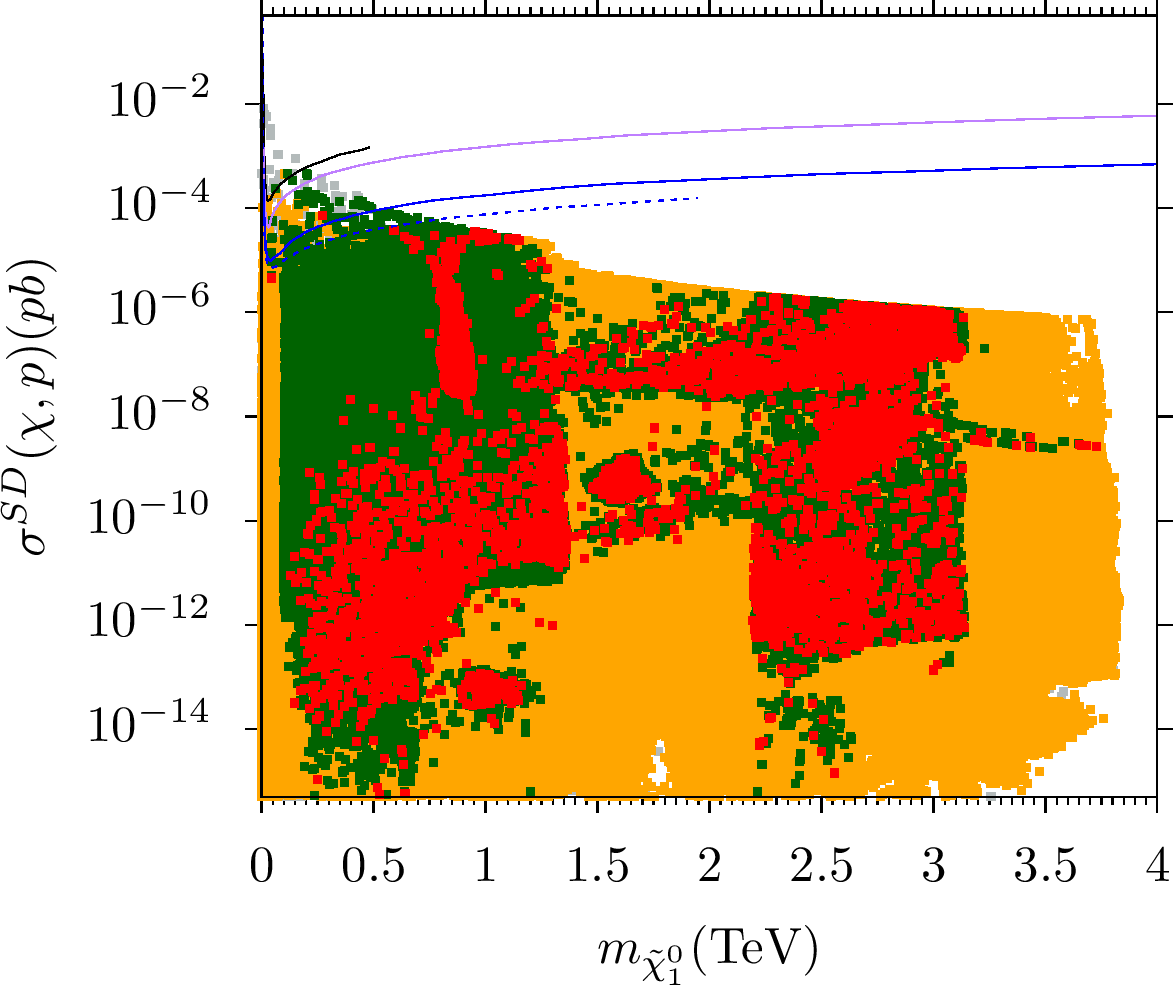}
    \caption{$m_{\tilde{\chi}^0_{1}}$ vs. $m_{\tilde {g}}$, $\sigma^{SI}(\chi,p)(pb)$ and $\sigma^{SD}(\chi,p)(pb)$: The left and right panels correspond to the $\mu > 0$ and $\mu < 0$ scenarios, respectively. Color coding is the same as that of Fig.\ref{fig1}. The solid black line represents the XENONnT \cite{XENON:2023cxc} and for the LUX-ZEPLIN, it is as follows: the Solid purple line represents LZ (2022), the solid blue line represents LZ (2024), and the dotted blue line represents the LZ-1000 day sensitivity~\cite{LZ:2018qzl, LZ:2022lsv,LZ:2024zvo}}.
		\label{fig7}
\end{figure*}

The top panels of Fig.~\ref{fig3} display the parameter space in the
$m_{\tilde t_1}$--$m_{\tilde{\chi}_1^0}$ plane, with the left (right) panel corresponds to the $\mu<0$ ($\mu>0$) scenario. The color coding follows the same convention as in Fig.~\ref{fig1}. The red points correspond to NLSP stop configurations that satisfy the $5\sigma$ Planck dark matter relic density constraint and are aligned along the diagonal black line, indicating an efficient stop-neutralino coannihilation mechanism. In the present analysis, viable NLSP stop solutions populate the mass range $0.9~\text{TeV} \lesssim m_{\tilde{t}_1} \lesssim 3.5~\text{TeV},$
significantly extending the parameter space explored in earlier studies. For instance, Ref.~\cite{Raza:2014upa} reported NLSP stop masses up to approximately $0.8~\text{TeV}$, while subsequent investigations of $b$--$\tau$ YU scenarios found an upper bound of about $3~\text{TeV}$~\cite{Baer:2012by,Raza:2018jnh}. We note that the relatively low population of viable points in certain regions of the $m_{\tilde{t}_1}$--$m_{\tilde{\chi}_1^0}$ plane is a consequence of the scan and can be populated if we do more focused scanning. The bottom panels of Fig.~\ref{fig3} show the mass splitting
$\Delta m_{\tilde t_1,\tilde{\chi}_1^0}$ as a function of the LSP neutralino mass,
$m_{\tilde{\chi}_1^0}$, with the left (right) panel corresponding to the $\mu<0$ ($\mu>0$) scenario. Compared to previous analyses, our results allow for substantially larger mass differences, reaching up to $\Delta m_{\tilde{t}_1,\tilde{\chi}_1^0} \sim 300~\text{GeV}$ for the red points. Importantly, these larger splittings occur in regimes where both the stop and neutralino masses are heavy, such that the relative mass difference remains small,$
\frac{\Delta m_{\text{NLSP,LSP}}}{m_{\text{LSP}}} \lesssim 10\%$,
thereby preserving the efficiency of the coannihilation process.
From a phenomenological perspective, larger mass splittings open up kinematically allowed two- and three-body decay channels, including $\tilde{t}_1 \to t\,\tilde{\chi}_1^0$, $\tilde{t}_1 \to W b \tilde{\chi}_1^0$, as well as four-body decays of the form $\tilde{t}_1 \to f f' b \tilde{\chi}_1^0$. On the other hand, for highly compressed spectra, they are not allowed kinematically, but the loop-induced two-body decay for the NLSP stop, $\tilde{t}_1 \to c\,\tilde{\chi}_1^0$ typically dominates over the four-body channel~\cite{Hikasa:1987db,Muhlleitner:2011ww}. Extensive experimental searches for NLSP stop scenarios have been performed at the LHC~\cite{ATLAS:2017drc,ATLAS:2017eoo,ATLAS:2017www,ATLAS:2017bfj,ATLAS:2019zrq,ATLAS:2021kxv,CMS:2022rqk,ATLAS:2020dsf,ATLAS:2020xzu,ATLAS:2021hza}. In most of these analyses, the stop mass was probed up to approximately $1.2~\text{TeV}$. By contrast, in our framework, the lightest NLSP stop mass consistent with all experimental, cosmological, and theoretical constraints is about $800~\text{GeV}$. Even in scenarios with small mass splittings where the decay $\tilde{t}_1 \to t\,\tilde{\chi}_1^0$ dominates, stop masses below $550~\text{GeV}$ have already been excluded~\cite{ATLAS:2021kxv}. Consequently, the NLSP stop solutions identified in this study lie well beyond current exclusion limits and represent promising targets for future searches at LHC Run-3 and beyond.

Figure~\ref{fig4} illustrates the stau-neutralino coannihilation region.  The top panels presents the parameter space in the
$m_{\tilde{\tau}_1}$--$m_{\tilde{\chi}_1^0}$ plane for $\mu<0$ (left) and($\mu>0$) (right). Color coding follows Fig.~\ref{fig1}. The upper panels highlight stau-neutralino coannihilation, while the lower panels explicitly display the mass splitting between the stau NLSP and neutralino LSP.
Our scan reveals that viable coannihilation solutions extend over a wide mass range,
$0.6~\text{TeV} \lesssim m_{\tilde{\tau}_1} \lesssim 3.8~\text{TeV}$,
with the stau and neutralino remaining nearly degenerate in mass. Such compressed spectra are essential for maintaining sufficiently large coannihilation rates in the early universe and are a generic feature of stau-assisted dark matter scenarios. The resulting parameter space is in good agreement with earlier analyses~\cite{Raza:2018jnh,Gomez:2020gav}, while significantly extending the reach toward heavier stau masses.
We note that our solutions are consistent with the CMS analysis based on $137~\mathrm{fb}^{-1}$ at $\sqrt{s}=13~\mathrm{TeV}$ ~\cite{CMS:2022rqk}. We anticipate that a portion of the parameter space presented here will be probed during LHC Run-3 and beyond.

Beyond the coannihilation channels discussed above, our scans identify viable chargino-neutralino coannihilation scenarios, as shown in Fig.~\ref{fig5}. The top panels display the parameter space in the
$m_{\tilde{\chi}_1^{\pm}}$--$m_{\tilde{\chi}_1^0}$ plane, while the bottom panels show the corresponding mass splitting,
$\lvert \Delta m_{\tilde{\chi}_1^{\pm},\tilde{\chi}_1^0} \rvert$, as a function of the chargino mass. Left (right) panels correspond to the $\mu<0$ ($\mu>0$) scenario. The red solutions span a broad mass range,
$0.2~\text{TeV} \lesssim m_{\tilde{\chi}_1^{\pm}} \lesssim 3.7~\text{TeV}$, and are in good agreement with earlier studies, such as Ref.~\cite{Gomez:2020gav}.
Recent LHC searches for electroweakinos place stringent constraints on chargino production. In particular, ATLAS has reported 95\% confidence level exclusion limits for slepton-mediated and SM boson-mediated decay channels 
$\tilde{\chi}_1^{\pm}\tilde{\chi}_1^{\mp}$ and $\tilde{\chi}_1^{\pm}\tilde{\chi}_2^0$ production~\cite{ATLAS:2021ilc}. These results indicate that, in scenarios where the chargino and neutralino are nearly mass-degenerate, chargino masses above approximately $300~\text{GeV}$ remain unconstrained. On the other hand, in regions of parameter space where slepton masses are heavier than those of the charginos, slepton-mediated decay channels are kinematically forbidden. In scenarios with a heavier NLSP chargino, we anticipate that such configurations may be probed in future LHC searches.

In addition to the various coannihilation mechanisms discussed above, our analysis also identifies viable Higgs-resonance solutions for neutralino DM. In this scenario, the relic abundance of the lightest neutralino is efficiently depleted through resonant annihilation into SM particles mediated by the CP-odd (even) Higgs bosons $A$ ($H$, $h$). Such resonance effects become particularly important when the kinematic condition $m_A \simeq 2 m_{\tilde{\chi}_1^0}$
is satisfied, leading to a significant enhancement of the annihilation cross section. As shown in Fig.~\ref{fig6}, our scans reveal a substantial region of parameter space consistent with this resonance condition, with the heavy Higgs masses obeying $m_A \simeq m_H$. These Higgs-funnel solutions successfully reproduce the observed DM relic density while remaining compatible with all imposed collider and cosmological constraints.
Current LHC searches for heavy neutral Higgs bosons decaying into $\tau^+\tau^-$ final states already impose meaningful restrictions on this scenario. In particular, CMS analyses exclude $m_A \lesssim 1.7~\text{TeV}$ for $\tan\beta \lesssim 30$~\cite{CMS:2022goy}. Furthermore, projected sensitivities indicate that for $\tan\beta \lesssim 10$, heavy Higgs masses up to approximately $1.0~\text{TeV}$, $1.1~\text{TeV}$, and $1.4~\text{TeV}$ may be exclude at the LHC Run-2, Run-3, and the High-Luminosity LHC, respectively~\cite{Baer:2022qqr,Baer:2022smj}. From our results, the Higgs-resonance solutions span a broad mass range,
$1.6~\text{TeV} \lesssim m_A \lesssim 3.7~\text{TeV}$,
indicating that while a portion of the parameter space has already been constrained by existing searches, a significant region remains unexplored. So some part of
the parameter space has already been explored by the LHC searches.

\begin{table}[h!]
	\centering
	\scalebox{0.7}{
		\begin{tabular}{|l|c|c|c|c|c|c|}
			\hline
			\hline
			&  Point 1 & Point 2  &  Point 3  & Point 4 & Point 5 & Point 6  \\
			\hline
			$m_{0}$        &  5189   & 19650   & 15710    &   4791 & 6167 & 18310  \\
			$M_{2} $          & 8489   & 8559    & 1529  & 9317   & -7477 & -8929    \\
			$M_{3}$         &  3205  & 1098   & 4680   & 4174  & 2611 & 911.5\\
			$A_{0}/m_{0}$  &  -0.7754 & 0.521 & 1.616   & -0.5652  & -2.344  & 0.1346 \\
			$\tan\beta$       & 49.94    & 52.41    & 20.08  &  48.07   & 10.59 & 43.81\\
			$m_{H_d}$         & 9414   & 9500   & 8606   &  9962  & 17270  & 6966\\
			$m_{H_u}$          & 2265   & 7451    & 1811  & 1625  & 4825 & 10730\\
			\hline
			$m_h$                & {\bf 125.48}   & {\bf 125.51}    & {\bf 125.51}   & {\bf 125.49}  & {\bf 125.48} & {\bf 125.52}\\
			$m_H$                & 6440.2  & 8944.9  & 15792 & 8060.6 & 18417 & 4969.6  \\
			$m_A$                 & 6398.1 & 8886.5  & 15688  & 8007.9  & 18297  & 4937.1    \\
			$m_{H^{\pm}}$         & 6441 & 8945.4  & 15792 & 8061.2 & 18418  & 4970.4\\
			\hline
			$m_{\tilde{\chi}^0_{1,2}}$
			&  {\color{red}2981}, 3504 & {\color{red}2676},  7463 & {\color{red} 1260}, 1273 & {\color{red} 3395},  3839& {\color{red} 1630}, 4712 & 2415, 7806   \\
			$m_{\tilde{\chi}^0_{3,4}}$
			& 3504, 7133 & 13218, 13218 & 14618, 14618  & 3840, 7806  & 4714, 6434 & 10355, 10355  \\
			
			$m_{\tilde{\chi}^{\pm}_{1,2}}$
			&  3387, 7068 &  7475, 13112& {\color{red} 1272}, 14630  & 3705, 7727  & 4735, 6444  & 7817, 10355 \\
			\hline
			$m_{\tilde{g}}$ &  6659   &{\color{red} 2774 }& 9951  & 8439 & 5728 & 2518         \\
			$m_{ \tilde{u}_{L,R}}$
			& 9114, 7510 & 20309, 19718 & 17437, 17403  & 10170, 8472  & 8998, 7102 & 19057, 18442   \\
			$m_{\tilde{t}_{1,2}}$
			& 4885, 6684 & 14484, 16765 & 12483, 15082 & 5896, 8016  & {\color{red}1644}, 7410 & 13005, 15709   \\
			\hline $m_{ \tilde{d}_{L,R}}$
			&9114, 7660 & 20310, 19734 & 17437, 17495  & 10170, 8618 & 8998, 7947  &19057, 18333    \\
			$m_{\tilde{b}_{1,2}}$
			& {\color{red}3034.4}, 6692 & 16627, 16894 & 14953, 16874  & 5236, 8033  & 7513, 7642  & 15604, 16046 \\
			\hline
			$m_{\tilde{\nu}_{e,\mu}}$
			& 7320   & 20325 & 15689& 7418 & 7289  & 19159     \\
			$m_{\tilde{\nu}_{\tau}}$   & 6454 & 18711 & 15278  & 6604 & 7184 & 17939 \\
			\hline
			$m_{ \tilde{e}_{L,R}}$
			& 7318, 6017 & 20321, 19776 & 15686, 15831  & 7417, 5868  & 7277, 7235 & 19154, 18297   \\
			$m_{\tilde{\tau}_{1,2}}$
			&  3461, 6445 & 16255, 18688 & 15008, 15278  & {\color{red}3399}, 6603  & 6974, 7170  &15610, 17919 \\
			\hline
			
			$\sigma_{SI}({\rm pb})$
			&  $3.28\times 10^{-11}$& 4.86$\times10^{-16}$ & 2.17$\times 10^{-12}$ & 9.65$\times 10^{-11}$ & $2.5\times 10^{-12}$ & $2.03\times 10^{-13}$\\

			$\sigma_{SD}({\rm pb})$
			&  1.1$\times 10^{-8}$ &1.3$\times 10^{-12}$ &2.52$\times 10^{-11}$   &1.22$\times 10^{-8}$  & $1.4 \times 10^{-10}$ & $8.93 \times 10^{-12} $\\
			$\Omega_{CDM}h^{2}$
			&  0.12538 & 0.12436 &0.12297  & 0.12108  & 0.12116 & 0.11773\\
			\hline
			\hline
		\end{tabular}
	}
	\caption{Fundamental parameters and the resulting sparticle mass spectrum are shown for $\mu<0$. All masses are given GeV. }
	\label{table1}
\end{table}

\begin{table}[h!]
	\centering
	\scalebox{0.7}{
		\begin{tabular}{|l|c|c|c|c|c|}
			\hline
			\hline
			&  Point 1 & Point 2  &  Point 3  & Point 4 & Point 5   \\
			\hline
			$m_{0}$        &  7207   & 6174   & 2531    & 4826 &  16980  \\
			$M_{2} $          & 8696   & 9749    & 3920  & 4894   & 7847    \\
			$M_{3}$         & 1148  & 1306   & 3071   & 2712  & 1010 \\
			$A_{0}/m_{0}$  &  -0.9271 & -0.7572 & -1.622   & -1.455  & 0.9791   \\
			$\tan\beta$       & 23.18    & 24.19    & 24.15  &  6.26  & 52.73 \\
			$m_{H_d}$         &9889   & 10870   & 14530   &  19490  & 7037 \\
			$m_{H_u}$          & 3708   & 3772    & 1650  & 3754  & 8138 \\
			\hline
			$m_h$                & {\bf 125.5}   & {\bf 125.5}    & {\bf 125.6}   & {\bf 125}  & {\bf 125.5} \\
			$m_H$                 & 11231 & 11745  & 14503  & 20571 & 4869  \\
			$m_A$                 & 11158 & 11668  & 14408  & 20436  & {\bf 4837}    \\
			$m_{H^{\pm}}$         & 11232 & 11745  & 14503 & 20571 & {\bf 4870} \\
			\hline
			$m_{\tilde{\chi}^0_{1,2}}$
			&  {\color{red}2671}, 4192 & {\color{red}854},  856 & {\color{red} 1626}, 3219 & {\color{red} 1859},  4133& {\color{red} 2429}, 6792   \\
			$m_{\tilde{\chi}^0_{3,4}}$
			& 4192, 7364 & 2984, 8240 & 3285, 3325  & 4786 ,4791  & 11583, 11583 \\
			
			$m_{\tilde{\chi}^{\pm}_{1,2}}$
			&  4263, 7322 & \textcolor{red} {882}, 8228& 3231, 3321  & 4151, 4796  & 6804, 11581  \\
			\hline
			$m_{\tilde{g}}$ &  \textcolor{red}{2749}   &3029& 6287  & 5741 & 2528         \\
			$m_{ \tilde{u}_{L,R}}$
			& 9142, 7376 & 8856,6449 & 6462, 5319  & 7511, 5858  & 17620, 17089    \\
			$m_{\tilde{t}_{1,2}}$
			& 3485, 7631 & 2147, 7433 & 3373, 5418 & \textcolor{red}{1876}, 6335  & 11839, 14210   \\
			\hline $m_{ \tilde{d}_{L,R}}$
			&9142, 7558 &8856, 6792 & 6463, 6194  & 7512, 7173 & 17620, 17040    \\
			$m_{\tilde{b}_{1,2}}$
			& 6913, 7700 & 5980, 7545 & 5432, 5480  & 6461, 7013  & 14098, 14518 \\
			\hline
			$m_{\tilde{\nu}_{e,\mu}}$
			&8922   & 8549 & 2768 & 4915 & 17670     \\
			$m_{\tilde{\nu}_{\tau}}$   & 8651 & 8291 & 1693  & 4856 & 15706  \\
			\hline
			$m_{ \tilde{e}_{L,R}}$
			& 8913, 7749 & 8536, 6951 & 2778, 4236  & 4910, 6517  & 17665, 17042    \\
			$m_{\tilde{\tau}_{1,2}}$
			&  7064, 8634 & 6195, 8236 & \textcolor{red}{1762}, 2961  & 4859, 6422  & 12593, 15685   \\
			\hline
			
			$\sigma_{SI}({\rm pb})$
			&  $8.09\times 10^{-11}$& 1.61$\times10^{-11}$ & 4.83$\times 10^{-12}$ & 2.11$\times 10^{-12}$ & $9.99\times 10^{-15}$ \\

			$\sigma_{SD}({\rm pb})$
			&  8.35$\times 10^{-10}$ &3.8$\times 10^{-8}$ &1.0$\times 10^{-9}$   &4.39$\times 10^{-11}$  & $2.74 \times 10^{-12}$ \\
			$\Omega_{CDM}h^{2}$
			&  0.11696 & 0.12233 &0.11849  & 0.11455  & 0.11406 \\
			\hline
			\hline
		\end{tabular}
	}
	\caption{Fundamental parameters and the resulting sparticle mass spectrum are shown for the $\mu>0$ scenario. All masses are given GeV. }
	\label{table2}
\end{table}

Finally, we examine the implications of current and future DM direct-detection experiments for the supersymmetric Pati- Salam model. In all viable coannihilation and Higgs-resonance scenarios discussed above, the LSP is predominantly bino-like, which plays a crucial role in determining the size of neutralino-nucleon scattering cross sections. Figure~\ref{fig7} presents the spin-independent (SI) neutralino-nucleon scattering cross section as a function of the LSP neutralino mass $m_{\tilde{\chi}_1^0}$ in the top panels, with the left (right) panel corresponding to the $\mu<0$ ($\mu>0$) scenario. The bottom panels display the corresponding spin-dependent (SD) scattering cross section for the same choices of $\mu$. In both the SI and SD panels, the solid black curve denotes the current XENONnT exclusion limit~\cite{XENON:2023cxc}. The LUX-ZEPLIN (LZ) constraints are shown by the solid purple line for LZ (2022), the solid blue line for the most recent LZ (2024) results, and the dotted blue line for the projected LZ 1000-day sensitivity~\cite{LZ:2018qzl,LZ:2022lsv,LZ:2024zvo}. From the $m_{\tilde{\chi}_1^0}$--$\sigma_{\rm SI}$ plane, we observe that the majority of the relic-density-saturating solutions (red points) lie below the current XENONnT and LZ (2022) exclusion limits. Only a small subset of red points, with neutralino masses in the range
$0.3~\text{TeV} \lesssim m_{\tilde{\chi}_1^0} \lesssim 3.5~\text{TeV}$,
has already been explored by the DD DM experiments. Importantly, a substantial fraction of the viable parameter space remains within the projected sensitivity of upcoming experiments: nearly half of the red points are accessible to the current LZ reach (solid blue curve), while approximately two-thirds of the solutions are expected to be probed by the LZ 1000-day exposure. The bottom panels, showing the $m_{\tilde{\chi}_1^0}$--$\sigma_{\rm SD}$ plane, indicate that all viable solutions comfortably satisfy existing spin-dependent bounds and remain consistent with the projected sensitivities of future direct-detection experiments. Overall, these results demonstrate that while the supersymmetric Pati-Salam model remains largely unconstrained by current direct-detection limits, a significant portion of the phenomenologically viable parameter space will be decisively tested by forthcoming LZ data, highlighting the strong complementarity between collider searches and DM experiments.

We conclude this section by presenting a set of representative benchmark points that exemplify the dominant DM mechanisms realized in the supersymmetric Pati-Salam model. In Table~\ref{table1}, several distinct coannihilation and resonance mechanisms are illustrated for the $\mu<0$ case. \textbf{Point~1} corresponds to a sbottom-neutralino coannihilation scenario, where the NLSP is a sbottom with mass $m_{\tilde{b}_1}=3.034~\text{TeV}$, nearly degenerate with a bino-dominated LSP neutralino of mass $m_{\tilde{\chi}_1^0}=2.891~\text{TeV}$, featuring small wino and Higgsino admixtures. \textbf{Point~2} realizes gluino-neutralino coannihilation, with an NLSP gluino at $m_{\tilde{g}}=2.774~\text{TeV}$ and a bino-like LSP at $m_{\tilde{\chi}_1^0}=2.676~\text{TeV}$. 
\textbf{Point~3} illustrates chargino-neutralino coannihilation, where the LSP neutralino, with a mixed bino-wino composition, has a mass of $m_{\tilde{\chi}_1^0}\simeq1.260~\text{TeV}$, while the lightest chargino lies at $m_{\tilde{\chi}_1^\pm}\simeq1.272~\text{TeV}$. \textbf{Point~4} corresponds to a stau-neutralino coannihilation scenario, characterized by an NLSP stau mass of $m_{\tilde{\tau}_1}\simeq3.399~\text{TeV}$ and a nearly degenerate bino-like LSP neutralino at $m_{\tilde{\chi}_1^0}\simeq3.395~\text{TeV}$. \textbf{Point~5} represents stop-neutralino coannihilation, with a light stop NLSP of mass $m_{\tilde{t}_1}\simeq1.644~\text{TeV}$ and a bino-dominated LSP neutralino at $m_{\tilde{\chi}_1^0}\simeq1.630~\text{TeV}$. 
Finally, \textbf{Point~6} exemplifies a Higgs funnel solution, in which the pseudoscalar and heavy scalar Higgs bosons are nearly degenerate, with $m_A=4.9371~\text{TeV}$ and $m_H=4.9704~\text{TeV}$. The bino-like LSP neutralino has a mass of $m_{\tilde{\chi}_1^0}\simeq2.415~\text{TeV}$, satisfying the resonance condition $m_{A,H}\approx2m_{\tilde{\chi}_1^0}$. This configuration yields an efficient annihilation channel through $A/H$ exchange while remaining consistent with current collider and DM searches. 

Table~\ref{table2} presents representative benchmark points for the $\mu>0$ scenario. \textbf{Point~1} realizes gluino-neutralino coannihilation, with a gluino NLSP at  $m_{\tilde{g}}\simeq2.749~\text{TeV}$ and a bino-like neutralino LSP at $m_{\tilde{\chi}_1^0}\simeq2.671~\text{TeV}$. \textbf{Point~2} demonstrates chargino-neutralino coannihilation, featuring a Higgsino-dominated neutralino LSP $m_{\tilde{\chi}_1^\pm}\simeq0.882~\text{TeV}$  nearly degenerate with its chargino partner $m_{\tilde{\chi}_1^0}\simeq0.854~\text{TeV}$.
\textbf{Point~3} corresponds to a stau-neutralino coannihilation scenario, featuring an NLSP stau mass of $m_{\tilde{\tau}_1}\simeq1.762~\text{TeV}$ and a bino-like LSP neutralino at $m_{\tilde{\chi}_1^0}\simeq1.626~\text{TeV}$. \textbf{Point~4} realizes stop-neutralino coannihilation, with a stop NLSP mass of $m_{\tilde{t}_1}\simeq1.876~\text{TeV}$ and a bino-dominated LSP neutralino at $m_{\tilde{\chi}_1^0}\simeq1.859~\text{TeV}$. Finally, \textbf{Point~5} exemplifies the Higgs funnel mechanism, where resonant annihilation occurs via nearly degenerate heavy Higgs bosons  $m_A=4.837~\text{TeV}$ and $m_H=4.870~\text{TeV}$, while the bino-like LSP neutralino has a mass of $m_{\tilde{\chi}_1^0}\simeq2.429~\text{TeV}$.

\section{Conclusion}
\label{sec:conc}
Driven by the growing agreement between the experimentally measured muon anomalous magnetic moment and its SM prediction, we reexamine phenomenological consequences of the MSSM, which is embedded in the supersymmetric $SU(4)_C \times SU(2)_L \times SU(2)_R$ Pati-Salam model. In contrast to earlier studies that predominantly favored a specific sign for the Higgsino mass parameter, our analysis systematically explores both $\mu > 0$, and $\mu < 0$ scenarios in light of current collider, cosmological, and DM constraints. Within this framework, we identify viable parameter space regions where the observed DM relic density is reproduced through multiple mechanisms: co-annihilations involving sbottom-neutralino, gluino-neutralino, stop-neutralino, stau-neutralino, and chargino-neutralino coannihilation, as well as resonant s-annihilation channel via the pseudoscalar Higgs boson. We demonstrate that all such scenarios are consistent with present bounds from LHC supersymmetry searches, the Planck~2018 DM relic density bound, and current limits from DD DM searches. Our results reveal characteristic mass spectra associated with these mechanisms. In particular, sbottom-neutralino coannihilation typically requires sbottom masses near $2.8~\text{TeV}$, while gluino-neutralino and stop-neutralino coannihilation scenarios allow gluino masses in the range $1$--$3~\text{TeV}$ and stop masses between $1$ and $3.5~\text{TeV}$. In coannihilation-dominated regions, the stau and chargino masses may reach values as high as $3.8~\text{TeV}$, whereas viable $A$ resonance solutions are realized for pseudoscalar Higgs masses spanning approximately $1.6$--$3.8~\text{TeV}$. We anticipate that a portion of the parameter space will be accessible to supersymmetry searches in LHC Run-3 and future runs.

\section*{Acknowledgements}
I.K. acknowledges support from Zhejiang Normal University through a postdoctoral fellowship under Grant No.~YS304224924.
TL is supported by the National Natural Science Foundation of China, by the Key Research Program of the Chinese Academy of Sciences, Grant No. XDPB15, by the Scientific Instrument Developing Project of the Chinese Academy of Sciences, Grant No. YJKYYQ20190049, and by the International Partnership Program of Chinese Academy of Sciences for Grand Challenges, Grant No. 112311KYSB20210012, and is supported in part by the National Key Research and Development Program of China Grant No. 2020YFC2201504, by the Projects No. 11875062, No. 11947302, No. 12047503, and No. 12275333


\end{document}